\documentclass[12pt]{article}

\newcommand{\blind}{1}

\RequirePackage{amsthm,amsmath,amsfonts,amssymb}
\allowdisplaybreaks[4]
\usepackage{enumitem}
\usepackage[round, authoryear]{natbib} 
\usepackage[colorlinks,citecolor=blue,urlcolor=blue,linkcolor=blue]{hyperref}
\usepackage{graphicx}
\usepackage{nicefrac}
\usepackage{booktabs}
\usepackage{multirow}
\usepackage{float}
\usepackage[font=footnotesize]{caption}
\usepackage{subcaption}
\usepackage[table,xcdraw]{xcolor}
\newcommand\red[1]{{\color{red}#1}}

\usepackage{setspace}

\usepackage{graphicx}
\graphicspath{{figures/}}

\theoremstyle{plain}

\newtheorem{theorem}{Theorem}[section]
\newtheorem{lemma}[theorem]{Lemma}

\newtheorem{assumption}{\textbf{Assumption}}

\newtheorem{remark}[theorem]{Remark}
\theoremstyle{remark}

\theoremstyle{definition}
\newtheorem{definition}[theorem]{Definition}
\newcommand{\bA}{\mathbf{A}}
\newcommand{\bB}{\mathbf{B}}
\newcommand{\bC}{\mathbf{C}}
\newcommand{\bD}{\mathbf{D}}

\newcommand{\bI}{\mathbf{I}}

\newcommand{\bK}{\mathbf{K}}

\newcommand{\bP}{\mathbf{P}}
\newcommand{\bQ}{\mathbf{Q}}
\newcommand{\bR}{\mathbf{R}}
\newcommand{\bS}{\mathbf{S}}
\newcommand{\bT}{\mathbf{T}}
\newcommand{\bU}{\mathbf{U}}
\newcommand{\bV}{\mathbf{V}}
\newcommand{\bW}{\mathbf{W}}
\newcommand{\bX}{\mathbf{X}}
\newcommand{\bY}{\mathbf{Y}}
\newcommand{\bZ}{\mathbf{Z}}

\newcommand{\be}{\mathbf{e}}
\newcommand{\bff}{\mathbf{f}}

\newcommand{\bx}{\mathbf{x}}
\newcommand{\by}{\mathbf{y}}

\newcommand{\bbR}{\mathbb{R}}
\newcommand{\bbN}{\mathbb{N}}

\newcommand{\bbC}{\mathbb{C}}
\newcommand{\calA}{\mathcal{A}}
\newcommand{\boldcalA}{\boldsymbol{\calA}}

\newcommand{\calN}{\mathcal{N}}

\newcommand{\calS}{\mathcal{S}}

\newcommand{\brho}{\boldsymbol{\rho}}
\newcommand{\btau}{\boldsymbol{\tau}}
\newcommand{\beps}{\boldsymbol{\varepsilon}}

\newcommand{\bSigma}{\boldsymbol{\Sigma}}
\newcommand{\bOmega}{\boldsymbol{\Omega}}
\newcommand{\bGamma}{\boldsymbol{\Gamma}}
\newcommand{\bPsi}{\boldsymbol{\Psi}}

\newcommand{\Prob}{\mathbb{P}}
\newcommand{\Expe}{\mathbb{E}}
\newcommand{\Var}{\mathrm{Var}}
\newcommand{\Cov}{\mathrm{Cov}}

\newcommand{\tr}{\mathrm{tr}}
\newcommand{\sign}{\mathrm{sign}}
\newcommand{\diag}{\mathrm{diag}}

\newcommand*{\dif}{\mathop{}\!\mathrm{d}}

\renewcommand{\bar}{\overline}
\usepackage{dsfont}
\newcommand{\indicator}{\mathds{1}}
\newcommand{\widesim}[2][1.5]{\mathrel{\overset{#2}{\scalebox{#1}[1]{$\sim$}}}}
\newcommand{\iidsim}{\widesim[2.33]{\mathrm{i.i.d.}}}	
\newcommand{\samedist}{\overset{\mathsf{d}}{=}}

\newcommand{\invGamma}{\mathsf{invGamma}}
\DeclareMathOperator*{\argmax}{arg\,max}
\DeclareMathOperator*{\argmin}{arg\,min}
\makeatletter
\renewcommand*{\top}{{\mathpalette\@transpose{}}}
\newcommand*{\@transpose}[2]{\raisebox{\depth}{$\m@th#1\intercal$}}
\makeatother
\newcommand{\SR}{\mathtt{SR}}
\newcommand{\NE}{\mathtt{NE}}
\newcommand{\BCV}{\mathtt{BCV}}
\newcommand{\ED}{\mathtt{ED}}
\newcommand{\ACT}{\mathtt{ACT}}

\newcommand{\MKTCR}{\mathtt{MKTCR}}

\begin{document}

\def\spacingset#1{\renewcommand{\baselinestretch}%
{#1}\small\normalsize} \spacingset{1}


\if1\blind
{
  \title{\bf Robust estimation for number of factors in high dimensional factor modeling via Spearman correlation matrix}
  \author{Jiaxin Qiu
  \\
    Department of Statistics and Actuarial Science\\The University of Hong Kong, Hong Kong, China\\
    {\small \texttt{qiujx@connect.hku.hk}}\\
    and\\
    Zeng Li\thanks{Zeng Li (\texttt{liz9@sustech.edu.cn}) is the corresponding author.} \\
    Department of Statistics and Data Science\\
    Southern University of Science and Technology, Shenzhen, China\\ 
    {\small \texttt{liz9@sustech.edu.cn}}\\
    and\\ 
    Jianfeng Yao\\
    School of Data Science\\
    The Chinese University of Hong Kong (Shenzhen), Shenzhen, China\\
    {\small \texttt{jeffyao@cuhk.edu.cn}}}
  \maketitle
} \fi

\if0\blind
{
  \bigskip
  \bigskip
  \bigskip
  \begin{center}
    {\LARGE\bf Robust estimation for number of factors in high dimensional factor modeling via Spearman correlation matrix}
\end{center}
  \medskip
} \fi

\bigskip
\begin{abstract}
Determining the number of factors in high-dimensional factor modeling is essential but challenging, especially when the data are heavy-tailed. In this paper, we introduce a new estimator based on the spectral properties of Spearman sample correlation matrix under the high-dimensional setting, where both dimension and sample size tend to infinity proportionally. Our estimator is robust against heavy tails in either the common factors or idiosyncratic errors. The consistency of our estimator is established under mild conditions. Numerical experiments demonstrate the superiority of our estimator compared to existing methods.

\end{abstract}

\noindent%
\vfill

\newpage
\spacingset{1.9} 


\section{Introduction}

Factor models are helpful tools for understanding the common dependence among high-dimensional outputs. They are widely used in data analysis in various areas like finance, genomics, and economics. Estimating the total number of factors is one of the most fundamental challenges when applying factor models in practice.
This paper focuses on the following factor model:
\begin{equation}\label{eq:factor-model}
	\by_i = \bB \bff_i + \bPsi\be_i,\qquad i\in[n]:= \{1,2,\ldots,n\},
\end{equation}
where $\{\by_i\}_{i=1}^n$ are the $p$-dimensional observation vectors, $\{\bff_i\}_{i=1}^n$ the $K$-dimensional latent common factor vectors, $\{\be_i\}_{i=1}^n$ the $p$-dimensional idiosyncratic error vectors, $\bB$ the $p\times K$ factor loading matrix, and $\bPsi$ a $p\times p$ diagonal matrix.
The objective of this paper is to estimate the number of common factors 
when the observed data are heavy-tailed.

There is a large literature on this estimation problem which can generally be categorized into two types of approaches.
The first type is based on information criteria. The seminal work \cite{bai2002determining}
proposed several information criteria, which were formulated in many different forms, through modifications of the Akaike information criterion (AIC) and the Bayesian information criterion (BIC). \cite{hallin2007determining} proposed an information criterion that utilized spectral density matrix estimation.  \cite{alessi2010improved} modified \cite{bai2002determining}'s criteria by tuning the penalty function to enhance their performance. \cite{kong2017number} employed similar ideas and put forth a local principal component analysis (PCA) approach to study a continuous-time factor model with time-varying factor loadings
using high-frequency data. \cite{li2017determining} used information criteria akin to those proposed by \cite{bai2002determining} for factor models when the number of factors increases with the cross-section size and time period. The first type of approach usually requires strong signals.
The second type of approach is based on the eigenvalue behavior of various types of covariance/correlation matrices. As for sample covariance matrices, \citet{Nadakuditi2008sample} proposed an estimator by exploiting the distribution properties of the moments of eigenvalues. \cite{ahn2013eigenvalue} proposed two estimators by utilizing the ratios of adjacent eigenvalues, namely the eigenvalue ratio (ER) estimator and the growth ratio (GR) estimator.  \cite{onatski2010determining,Onatski2012asymptotic} proposed an alternative edge distribution (ED) estimator based on the maximum differences between consecutive eigenvalues instead of their ratios. \citet{Owen2016BCV} introduced an estimator that utilizes the bi-cross-validation
(BCV) technique from \citet{Owen2009BCV}. This estimator is based on the theoretical results concerning the spiked sample covariance matrix.
For lagged sample autocovariance matrices, \cite{lam2012factor} developed a ratio-based estimator for factor modeling of multivariate time series. This estimator was further extended by \cite{li2017identifying} to accommodate weak factors. 
As for correlation matrices, \cite{fan2020estimating} proposed a tuning-free and scale-invariant adjusted correlation thresholding method.
This approach has been further extended to a time series tensor factor model in \cite{lam2021rank} and \cite{chen2024rank}.

\begin{figure}[htbp]
	\centering
	\begin{subfigure}{.4\textwidth}
		\includegraphics[width=\textwidth]{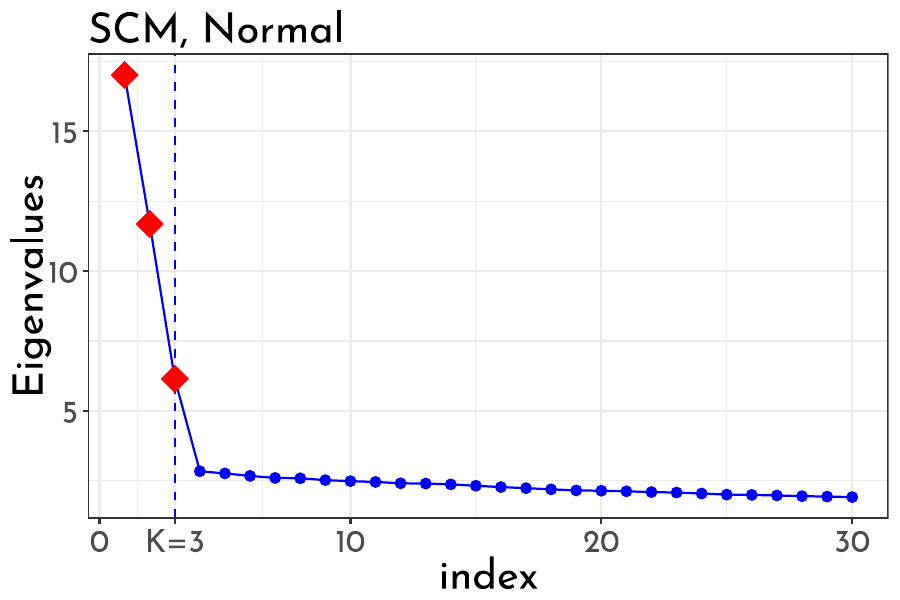}
		\caption{Sample covariance matrix}
		\label{fig:SCM_scatter_Normal}
	\end{subfigure}%
        \qquad
	\begin{subfigure}{.4\textwidth}
		\includegraphics[width=\textwidth]{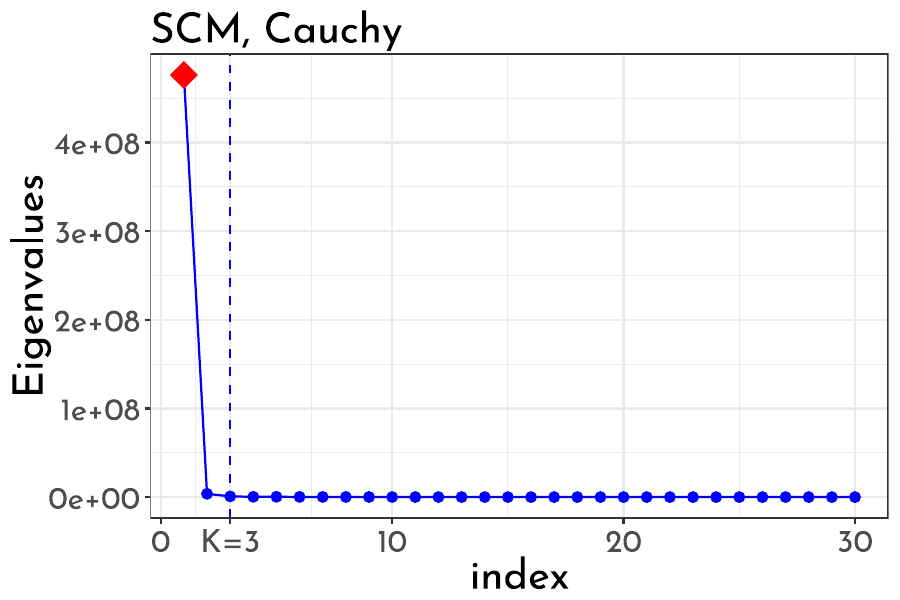}
		\caption{Sample covariance matrix}
		\label{fig:SCM_scatter_Cauchy}
	\end{subfigure}%
 
	\begin{subfigure}{.4\textwidth}
		\includegraphics[width=\textwidth]{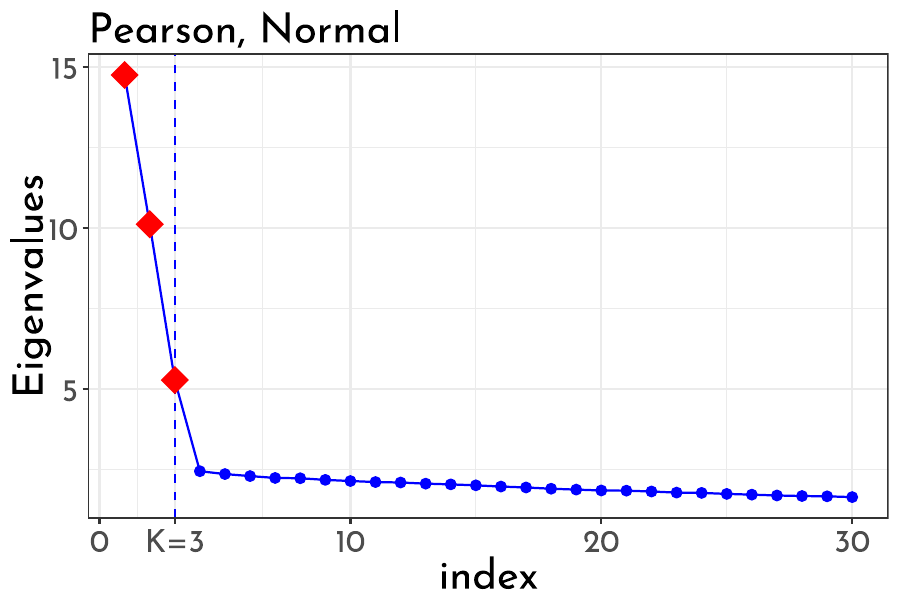}
		\caption{Pearson  correlation matrix}
		\label{fig:Pearson_scatter_Normal}
	\end{subfigure}%
        \qquad
        \begin{subfigure}{.4\textwidth}
		\includegraphics[width=\textwidth]{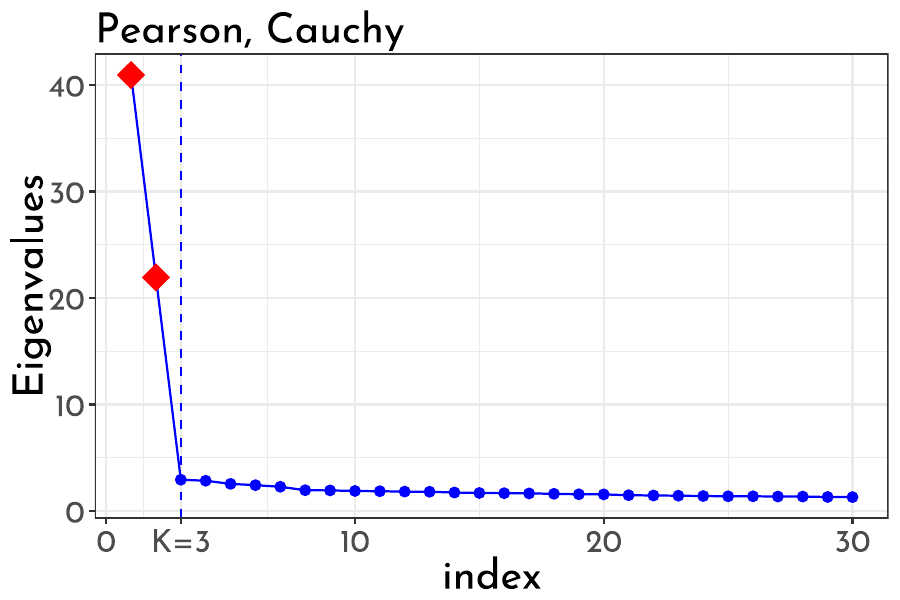}
		\caption{Pearson  correlation matrix}
		\label{fig:Pearson_scatter_Cauchy}
	\end{subfigure}%

        \begin{subfigure}{.4\textwidth}
		\includegraphics[width=\textwidth]{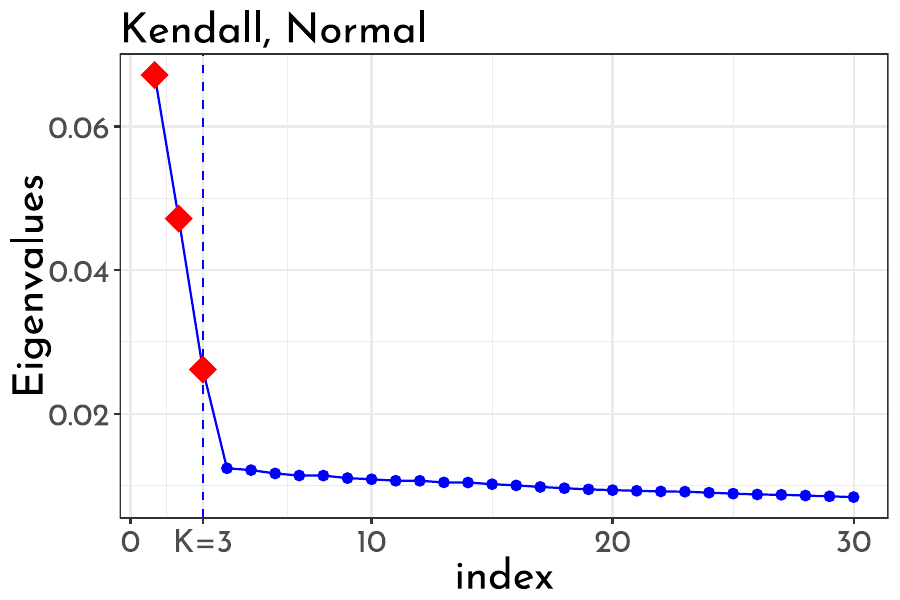}
		\caption{Multivariate Kendall's tau matrix}
		\label{fig:MKen_scatter_Normal}
	\end{subfigure}%
        \qquad
	\begin{subfigure}{.4\textwidth}
		\includegraphics[width=\textwidth]{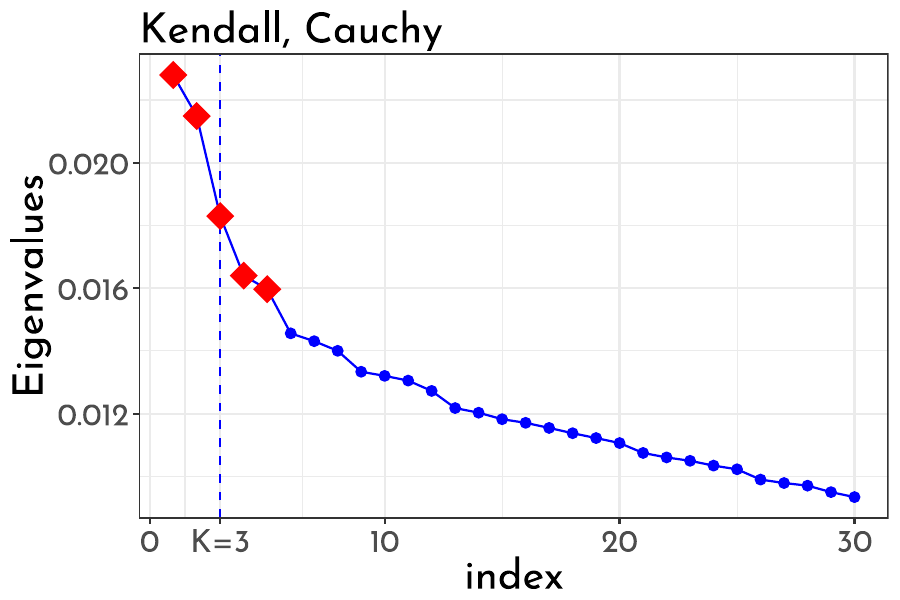}
		\caption{Multivariate Kendall's tau matrix}
		\label{fig:MKen_scatter_Cauchy}
	\end{subfigure}%
 
	\begin{subfigure}{.4\textwidth}
		\includegraphics[width=\textwidth]{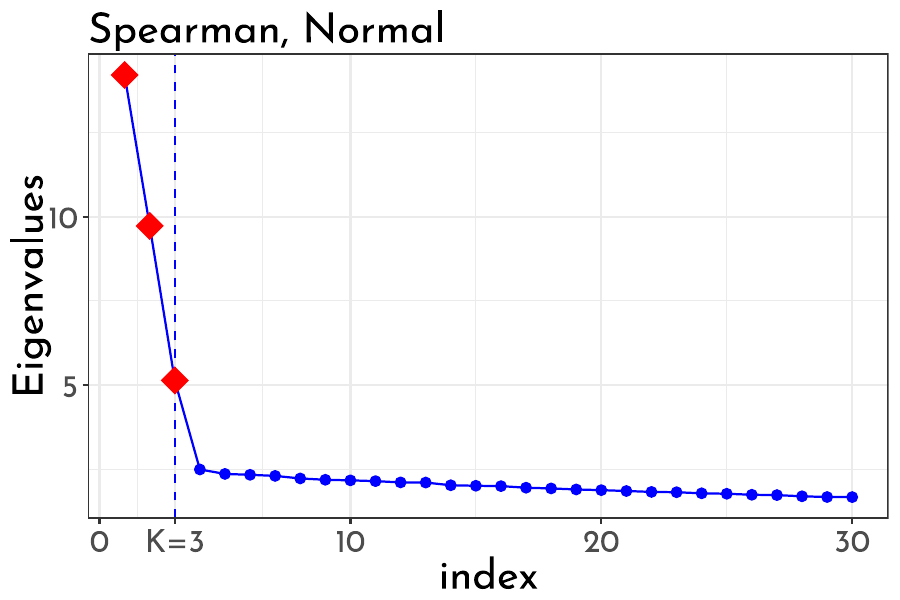}
		\caption{Spearman correlation matrix}
		\label{fig:Spearman_scatter_Normal}
	\end{subfigure}%
        \qquad
        \begin{subfigure}{.4\textwidth}
		\includegraphics[width=\textwidth]{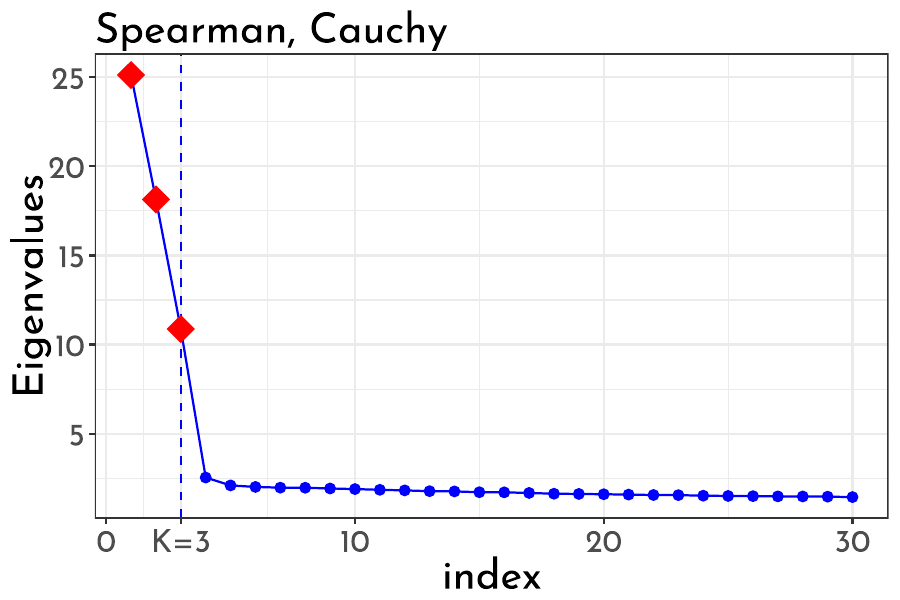}
		\caption{Spearman correlation matrix}
		\label{fig:Spearman_scatter_Cauchy}
	\end{subfigure}%
	\caption{Scatter plots of the first $30$ eigenvalues of the sample covariance matrix (SCM), Pearson  correlation matrix, Spearman correlation matrix, and multivariate Kendall's tau matrix. Data are generated following the factor model \eqref{eq:factor-model} with $K=3$. The factors and idiosyncratic errors are drawn independently from standard \textbf{Normal} distribution (\textbf{Left panel}: Figures \subref{fig:SCM_scatter_Normal}, \subref{fig:Pearson_scatter_Normal}, \subref{fig:MKen_scatter_Normal}, \subref{fig:Spearman_scatter_Normal}) or standard \textbf{Cauchy} distribution (\textbf{Right panel}: Figures \subref{fig:SCM_scatter_Cauchy}, \subref{fig:Pearson_scatter_Cauchy}, \subref{fig:MKen_scatter_Cauchy}, \subref{fig:Spearman_scatter_Cauchy}). The symbol ``\red{$\blacklozenge$}'' represents the spiked eigenvalues. Further details regarding the matrices $\bB$ and $\bPsi$ can be found in the case (C1) in Section \ref{sec:simu}.}
  \label{fig:scatter_eigvals}
\end{figure}

The aforementioned methods have been proved to be inadequate when dealing with heavy-tailed data, and would mostly result in biased or inconsistent estimators. 
Heavy-tailed data are common in various real-world applications. For instance, prices of stock returns often exhibit heavy tails due to the occurrence of extreme events in the market. However, little literature has focused on estimating the number of factors in the context of heavy-tailed data. 
Assuming a jointly elliptical distribution for both common factors and idiosyncratic errors (as discussed in \cite{fan2018large}), \cite{yu2019robust} proposed two estimators utilizing the sample multivariate Kendall's tau matrix. \cite{he2022matrix} further extended it to the matrix factor model. \cite{he2022large} recovered factor loadings and scores by performing PCA on the multivariate Kendall's tau matrix. It is worth mentioning that \cite{yu2019robust}'s method requires that $\|\bB^{\top}\bB/p-\bSigma_B\|_2\to 0$, where $\bSigma_B$ is a $K\times K$ positive definite matrix with bounded and distinct eigenvalues (see their Assumptions 2.3). The factor model is considered to have a strong factor structure \citep{bai2002determining} when both $\bB^{\top}\bB/p$ and $\sum_{i=1}^n\bff_i\bff_i^{\top}/n$ converge to positive definite matrices. In this paper, we consider the weak loading scenario by assuming $\bB^{\top}\bB=\bI_K$, which is a commonly used identifiability condition in the literature on factor models (see, for example, \citet{bai2012statistical}). Moreover, we address a more challenging scenario where both the factors and idiosyncratic errors may be heavy-tailed, potentially leading to the non-existence of the limit of $\sum_{i=1}^n\bff_i\bff_i^{\top}/n$. 
We propose an estimator based on \emph{Spearman correlation matrix} \citep{spearman1961proof} which shows significant improvements over existing methods. Here, we use a toy example to demonstrate the robustness of Spearman correlation matrix. 
Data are generated following factor model \eqref{eq:factor-model} with $K=3$. The factors and idiosyncratic errors follow either standard normal distribution or standard Cauchy distribution. 
As shown in Figure \ref{fig:scatter_eigvals}, when the common factors and the idiosyncratic noise are light-tailed, all four sample covariance/correlation matrices have three spiked eigenvalues, and all factors can be detected. When the data distribution is heavy-tailed, only  our method can clearly identify all three factors.

The Spearman correlation matrix is defined as the Pearson correlation matrix of the ranks of the data. It is a valuable tool when dealing with heavy-tailed data. However, the nonlinear structure of rank-based correlation brings significant difficulties when analyzing its eigenvalue behavior. To address this, we need to resort to tools in random matrix theory (RMT). Unfortunately, most existing work in RMT focuses on very restrictive settings where data has independent components. \cite{bai2008large} showed that its limiting spectral distribution (LSD) is the well-known Mar\v{c}enko-Pastur law. \cite{bao2015spectral} established the central limiting theorem (CLT) for its linear spectral statistics (LSS). \cite{bao2019tracy-spearman} showed that the Tracy-Widom law holds for its largest eigenvalues. To the best of our knowledge, the first investigation of the Spearman sample correlation matrix for general dependent data was conducted very recently by \cite{wu2022limiting}, which derived its LSD under the non-paranormal distribution proposed by \cite{JMLR:v10:liu09a}. Many other spectral properties, including the extreme eigenvalues, CLT for LSS, and spiked eigenvalues for dependent data, still remain open.
Our work is the first to investigate the eigenvalue behavior of the Spearman sample correlation matrix under spike models, and successfully apply the theories to identify the number of factors in high-dimensional factor modeling for heavy-tailed data.

To summarize, the main contributions of this paper are two-fold. First, we propose a new estimator based on the Spearman sample correlation matrix for the number of common factors in the high-dimensional factor model \eqref{eq:factor-model}. This estimator is distribution-free and capable of estimating the number of factors even when the data is heavy-tailed.
Second, we provide a theoretical explanation of the phase-transition phenomenon for the top eigenvalues of the Spearman sample correlation matrix under the spike model. 
From a technical point of view, we investigate this phase-transition theory by establishing the universality of the asymptotic law of a low-dimensional random matrix (see Lemma \ref{lem:G4MT} and Remark \ref{rmk:phase-transition-proof} for more details), and our method does not require the commonly used independent component structure.

Before moving forward, let us introduce some notations that will be used throughout this paper. We use $[n]$ to denote the set $\{1,2,\ldots,n\}$. We adopt the convention of using regular letters for scalars, and bold-face letters for vectors or matrices.
For any matrix $\bA$, we denote its $(i,j)$-th entry by $A_{ij}$, its transpose by $\bA^{\top}$, its trace by $\tr(\bA)$, its $j$-th largest eigenvalue by $\lambda_j(\bA)$ (when the eigenvalues of $\bA$ are real), its spectral norm by $\|\bA\|_2=\sqrt{\lambda_1(\bA\bA^{\top})}$, and its element-wise maximum norm by $\|\bA\|_{\max}=\max\limits_{i,j}|A_{ij}|$. 
We use $\diag(\bA)$ to denote the diagonal matrix of $\bA$ (replacing all off-diagonal entries with zero).
For a sequence of random variables $\{X_n\}_{n=1}^{\infty}$ and a corresponding set of nonnegative real numbers $\{a_n\}_{n=1}^{\infty}$, we write $X_n=O_P(a_n)$ if $X_n/a_n=O_P(1)$ (bounded in probability), and we write $X_n=o_P(a_n)$ if $X_n/a_n\to 0$ in probability. For any univariate function $f$, we denote $f(\bA)=[f(A_{ij})]$ as a matrix with $f$ applied on each entry of $\bA$.
Throughout this paper, $C$ stands for some positive constant whose value is not important and may change from line to
line. The notation ``$i_1\neq i_2\neq\cdots\neq i_m$'' indicates that the $m$ indices $\{i_{\ell}\}_{\ell=1}^m$ are pairwise different. All limits are for $n\to \infty$, unless explicitly stated otherwise.

The rest of this article is organized as follows.  Section \ref{sec:main-result} proposes a new estimator for the number of common factors in the factor model \eqref{eq:factor-model}. The consistency of our estimator is proven based on the spectral properties of the Spearman sample correlation matrix. Section \ref{sec:simu} offers comprehensive simulation experiments, comparing our estimator with others. In Section \ref{sec:real-data-example}, we evaluate the performance of the proposed estimator on a real dataset. A brief discussion is given in Section \ref{sec:discussion}. Section \ref{sec:lemma_proof_theorem} presents some important technical lemmas and proofs. Auxiliary lemmas and technical proofs are relegated to the supplementary material.

\addtolength{\textheight}{.5in}%

\section{Main results}\label{sec:main-result}

\subsection{Spearman correlation matrix}

For $p$-dimensional i.i.d. data sample $\{\by_i\}_{i=1}^n$, we denote the ranks of the data as follows:
\[
\bY_n = \begin{pmatrix}
    \by_1^{\top}\\
    \vdots\\
    \by_n^{\top}
\end{pmatrix}
=\underbrace{\begin{pmatrix}
    y_{11} & \cdots & y_{1p}\\
    \vdots & \ddots & \vdots\\
    y_{n1} & \cdots & y_{np}
\end{pmatrix}}_{\text{raw data matrix}}
\implies 
\underbrace{\begin{pmatrix}
    r_{11} & \cdots & r_{1p}\\
    \vdots & \ddots & \vdots\\
    r_{n1} & \cdots & r_{np}
\end{pmatrix}}_{\text{ranks matrix}},
\]
where $r_{ij}=\sum_{\ell=1}^n \indicator\{y_{\ell j}\leqslant y_{ij}\}$ is the rank of $y_{ij}$ among $\{y_{\ell j}\}_{\ell=1}^n$, and $\indicator\{\cdot\}$ denotes the indicator function. The Spearman correlation matrix of the raw data matrix $\bY_n$ is the Pearson correlation matrix of the ranks matrix. Define the normalized ranks matrix
\begin{equation}\label{eq:rank_matrix}
	\bR = \sqrt{\frac{12}{n^2-1}}\Bigl(r_{ij}-\frac{n+1}{2}\Bigr)_{n\times p},
\end{equation}
and let $\bR_i^{\top}$ be the $i$-th row of the matrix $\bR$. The Spearman sample correlation matrix of $\bY_n$ is
\begin{equation}\label{eq:rho_matrix}
	\brho_n = \frac{1}{n}\bR^{\top}\bR= \frac{1}{n} \sum_{i=1}^n \bR_i\bR_i^{\top}.
\end{equation}
The empirical spectral distribution (ESD) of $\brho_n$ is referred to as a random measure
$
	F^{\brho_n} = p^{-1}\sum_{j = 1}^{p} \delta_{\lambda_j(\brho_n)},
$
where $\delta_{\lambda_j(\brho_n)}$ is the Dirac mass at the point $\lambda_j(\brho_n)$. The limit of $F^{\brho_n}$ is called limiting spectral distribution (LSD). Under the assumption that the components of $\by_i$ are i.i.d., \cite{bai2008large} proved that the LSD of $\brho_n$ is the well-known Mar\v{c}enko-Pastur distribution. Recently, \cite{wu2022limiting} extended this result to the non-paranormal distribution. In this study, we further extend their findings to encompass the scale mixture of normal distributions (see Definition \ref{def:SMN}), as stated in Lemma \ref{lem:rho_W_LSD}.

Throughout this paper, we assume that both common factors and  idiosyncratic errors follow continuous distributions. Therefore, with probability one, there are no ties among $\{y_{ij}, i\in [n]\}$ for each $j$. For any $j\in[p]$ and $i, \ell\in[n]$ with $i\neq \ell$, we have $\indicator\{y_{\ell j}\leqslant y_{i j}\}=\frac{1}{2}+\frac{1}{2}\sign( y_{i j} -  y_{\ell j})$, where $\sign(\cdot)$ denotes the sign function. Hence, we have
\begin{equation}\label{eq:rank_sign}
	r_{ij} -\frac{n+1}{2}= 1 + \frac{1}{2} \sum_{i\neq \ell} \bigl\{1+\sign( y_{ij} -  y_{\ell j})\bigr\} - \frac{n+1}{2} = \frac{1}{2}\sum_{i\neq \ell} \sign( y_{ij} -  y_{\ell j}).
\end{equation}
For two sample vectors $\by_i$ and $\by_{\ell}$, we define the sign vector
\[
\bA_{i\ell} = \sign(\by_i-\by_{\ell})=\bigl(\sign(y_{i 1}-y_{\ell 1}), \ldots, \sign(y_{i p}-y_{\ell p})\bigr)^{\top}.
\] 
Then, from \eqref{eq:rank_matrix} and \eqref{eq:rank_sign}, we can rewrite the Spearman sample correlation matrix \eqref{eq:rho_matrix} as
\begin{equation*}
\brho_n=\frac{3}{n(n^2-1)}\sum_{i=1}^n\sum_{\ell_1, \ell_2\neq i} \bA_{i\ell_1}\bA_{i\ell_2}^{\top}.
\end{equation*}
The application of sign transformations to the data introduces an intractable nonlinear correlation structure. 
To address this challenge, we utilize \emph{Hoeffding's decomposition} \citep{hoeffding1948Aclass} to handle the nonlinear correlation within $\bA_{i\ell}$. By employing this decomposition, we can identify the dominant term of $\brho_n$.
Let $\bA_i=\Expe (\bA_{i\ell}\mid\by_i)$ with $i\neq \ell$, Hoeffding's decomposition of $\bA_{i\ell}$ can be expressed as follows:
\begin{equation}\label{eq:hoeffding-decom}
\bA_{i\ell} = \bA_i-\bA_{\ell}+\beps_{i\ell},
\end{equation}
where $\beps_{i\ell}:=\bA_{i\ell}-\bA_i+\bA_{\ell}$. Note that $\Expe \bA_{i\ell}=\Expe \bA_i = \boldsymbol{0}$, and the covariance matrix of $\bA_i$ is $\Expe (\bA_i\bA_i^{\top}).$
With Hoeffding's decomposition defined in \eqref{eq:hoeffding-decom}, we have
\begin{align*}
    \brho_n
    & = \frac{3}{n(n^2-1)}\Biggl\{\sum_{i=1}^n\sum_{\ell_1, \ell_2\neq i} (\bA_i-\bA_{\ell_1})(\bA_i-\bA_{\ell_2})^{\top}+ \sum_{i=1}^n\sum_{\ell_1, \ell_2\neq i} (\bA_i-\bA_{\ell_1})\beps_{i\ell_2}^{\top}\\
    &\qquad \qquad\qquad\qquad
     + \sum_{i=1}^n\sum_{\ell_1, \ell_2\neq i} \beps_{i\ell_1}(\bA_i-\bA_{\ell_2})^{\top}  + \sum_{i=1}^n\sum_{\ell_1, \ell_2\neq i} \beps_{i\ell_1}\beps_{i\ell_2}^{\top}\Biggr\}.
\end{align*}
It will be shown that the cross-terms in the above identity are negligible (see Lemma \ref{lem:rho_W_spectral_norm} and its proof in Section S2.1 in the supplementary material).
We can then focus on the first term, 
\begin{align}
    &\;\frac{3}{n(n^2-1)}\sum_{i=1}^n\sum_{\ell_1, \ell_2\neq i} (\bA_i-\bA_{\ell_1})(\bA_i-\bA_{\ell_2})^{\top}\nonumber\\
    = &\;\frac{n-2}{n+1}\cdot\frac{3}{n}\sum_{i=1}^n \biggl\{\frac{1}{(n-1)(n-2)}\sum_{\ell_1\neq \ell_2\neq i} (\bA_i-\bA_{\ell_1})(\bA_i-\bA_{\ell_2})^{\top} \biggr\} +\frac{3}{n+1}\btau_n,\label{eq:rho-hoeffding-decom-2}
\end{align} 
where 
$
\btau_n := \frac{1}{n(n-1)}\sum_{i=1}^n\sum_{\ell\neq i} (\bA_i-\bA_{\ell})(\bA_i-\bA_{\ell})^{\top}
$
is the sample marginal Kendall's tau correlation matrix \citep{bandeira2017marvcenko, li2023eigenvalues}.
The second term, $3\btau_n/(n+1)$, and the cross-terms in \eqref{eq:rho-hoeffding-decom-2} are negligible (see Lemma \ref{lem:rho_W_spectral_norm} and its proof in Section S2.1 in the supplementary material).
Hence, the leading order term of $\brho_n$ is
\begin{equation}\label{eq:W}
	\bW_n = \frac{3}{n}\sum_{i=1}^n \bA_i\bA_i^{\top}.
\end{equation}
Through direct calculations, it can be demonstrated that the difference between the expected values of $\bW_n$ and $\brho_n$ is of the order $O_P(n^{-1})$. To be more precise, we have $\Expe \brho_n = \frac{n}{n+1}\Expe \bW_n + \frac{3}{n+1}\Expe \beps_{12}\beps_{12}^{\top}$.

Under certain assumptions, we can show that the spectrum of $\brho_n$ can be approximated by that of $\bW_n$ (see Lemma \ref{lem:rho_W_spectral_norm}). This allows us to study the spectral properties of $\brho_n$ via those of $\bW_n$.
Under the scale mixture of normals framework defined in Section \ref{sec:SMN}, we find that the population covariance matrix $\bSigma_{\rho}=\Expe \bW_n$ has a finite-rank perturbation structure with $K$ spiked eigenvalues (see Lemma \ref{lem:Sigma_finite_rank_perturbation}). Thus naturally $\bW_n$ has $K$ relatively large eigenvalues too. Subsequently, we can estimate the number of factors based on the top eigenvalues of $\bW_n$ or $\brho_n$. By establishing the phase-transition theory of the spiked eigenvalues of $\bW_n$, the consistency of the new estimator follows.

\subsection{Phase transition theory}\label{sec:SMN}

From the perspective of RMT, the spectrum of $\bW_n$ relies on the structure of $\bSigma_{\rho}=\Expe \bW_n$. 
Although from factor model \eqref{eq:factor-model}, it is clear that $\bSigma_y:=\Cov(\by_i)=\bB\bB^{\top}+\bPsi$ has a finite-rank-$K$ perturbation structure, the relationship between $\bSigma_{\rho}$ and $\bSigma_y$ is unclear. The structure of $\bSigma_{\rho}$ changes for different distributions of $\by_i$. Therefore, extra distribution assumption of $\by_i$ is needed to maintain the finite-rank perturbation structure of $\bSigma_{\rho}$.
Specifically, we assume that both the common factors and the idiosyncratic errors follow a \emph{scale mixture of normal distributions}, defined as follows:

\begin{definition}[Scale mixture of normals, \cite{andrews1974scale}]\label{def:SMN}
	A $p$-dimensional random vector $\bX=(X_1,\ldots,X_p)^{\top}$ follows a \emph{scale mixture of normal distributions} (SMN) if $\bX$ has the stochastic representation
    $
	    \bX \samedist \sqrt{W}\bZ,
    $
	where $W$ is a scalar-valued random variable with positive support, and $\bZ$ follows $p$-dimensional normal distribution $\calN_p(\boldsymbol{0},\bSigma)$ independent of $W$, $\bSigma$ is a positive semi-definite matrix. The notation ``$X\samedist Y$'' means $X$ and $Y$ have the same distribution. 
\end{definition}

Our motivation for using this scale mixture of normals is two-fold. First, the scale mixture of normals contains heavy-tailed distributions, such as Student's $t$ distribution. If $W$ follows the inverse Gamma distribution $\invGamma(\nu/2,\nu/2)$ with probability density function
$
g_W(w) = \frac{(\nu/2)^{\nu/2}}{\Gamma(\nu/2)} w^{-(\nu/2+1)}\exp\{-\nu/(2w)\},
$
then $\bX$ follows the $p$-dimentional Student's $t$ distribution $t_{\nu}(\boldsymbol{0},\bSigma)$ with location parameter $\boldsymbol{0}$, scale matrix $\bSigma$, and degrees of freedom $\nu$, the probability density function of which is
$
f_{\bX}(\bx)=\frac{\Gamma((\nu+p)/2)}{\Gamma(\nu/2)\nu^{p/2}\pi^{p/2}|\bSigma|^{\nicefrac{1}{2}}}(1+\frac{1}{\nu}\bx^{\top}\bSigma^{-1}\bx)^{-\nicefrac{(\nu+p)}{2}}.
$
A number of well-known distributions can be written as scale mixtures of normals. We refer the readers to Section 2 of \cite{heinen2020spearman} for more examples.
The second motivation is for technical advantage. 
From the fact that $\bX\mid (W=w) \sim\calN_p(\boldsymbol{0},w\bSigma)$, we can relate the Spearman sample correlation matrix of $\bX$ to the scale matrix $\bSigma$ using Grothendieck's identity (see Lemma S1.5 in the supplementary material). 
\cite{heinen2020spearman} derived an explicit expression for Spearman correlation of bivariate scale mixture of normals. We extend this result to a more complicated bivariate population (see Lemma S1.6 in the supplementary material)
and utilize it to examine the structure of $\bSigma_{\rho}$ (see Lemma \ref{lem:Sigma_finite_rank_perturbation}). This direct connection to the scale matrix $\bSigma_{\rho}$ is a fundamental step in the analysis of our proposed estimator.

Furthermore, we need the following assumptions:
\begin{assumption}\label{assump-p-n}%
As $n\to \infty$, $p=p(n)\to\infty$ and $p/n=c_n\to c \in (0,\infty)$. 
\end{assumption}


\begin{assumption}\label{assump-SMN-different-W}
  All pairs $\{(\bff_i^{\top},\be_i^{\top})^{\top}\}_{i=1}^n$ are i.i.d., and $\bff_i$ is independent of $\be_i$, and both of them follow the scale mixture of normal distributions.
	Suppose that $w_f$ and $w_e$ are two independent random variables with positive support. The common factor $\bff_i$ has a stochastic representation $\bff_i \samedist \sqrt{w_i^f}\bx_i$, where $w_i^f$ is an independent copy of $w_f$, and $\bx_i$ follows $K$-dimensional standard normal distribution. The idiosyncratic error $\be_i$ has a stochastic representation
  $\be_i\samedist (\sqrt{w_{i1}^e}z_{i1},\ldots,\sqrt{w_{ip}^e}z_{ip})^{\top}$, where $\{w_{ij}^e\}_{j=1}^p$ are i.i.d. copies of $w_e$, and $(z_{i1},\ldots,z_{ip})^{\top}$ follows $p$-dimensional standard normal distribution.
\end{assumption}

\begin{assumption}\label{assump-loading}%
	The loading matrix $\bB$ is normalized by the constraint $\bB^{\top}\bB=\bI_K$. All the entries of $\bB$ are of order $O(p^{-\nicefrac{1}{2}})$. 
\end{assumption}

\begin{assumption}\label{assump-psi}%
	The matrix $\bPsi$ is diagonal with entries of order $O(1)$.
\end{assumption}

Assumption \ref{assump-p-n} is common in the RMT literature. 
Assumption \ref{assump-SMN-different-W}
pertains to the distribution of the common factors and the idiosyncratic errors, and allows them to be heavy-tailed. This assumption is crucial for examining the structure of the population covariance matrix $\bSigma_{\rho}$.
The constraint $\bB^{\top}\bB=\bI_K$ in \ref{assump-loading} is a commonly used identifiability condition, see \cite{bai2012statistical}. It follows that by singular value decomposition, we can represent $\bB$ as $\bU\bV^{\top}$, where $\bU\in\bbR^{p\times K}$ and $\bV\in\bbR^{K\times K}$, and both have orthonormal columns. The column vectors of $\bU$ and $\bV$ are unit vectors in $\bbR^p$ and $\bbR^K$, respectively. Thus, the condition $B_{ij} = O(p^{-\nicefrac{1}{2}})$ in \ref{assump-loading}, for any $i\in[p]$ and  $j\in [K]$, is not overly restrictive. This condition  facilitates our technical proofs.
Assumption \ref{assump-psi} is standard in the factor models literature. 

In what follows, we develop some important spectral properties of the Spearman sample correlation matrix. First, we show that the spectrum of $\brho_n$ can be approximated by that of $\bW_n$, as stated in the following lemma.
\begin{lemma}\label{lem:rho_W_spectral_norm} 
Under Assumptions \ref{assump-p-n}  --  \ref{assump-psi}, for any $j\in[p]$, $|\lambda_j(\brho_n)-\lambda_j(\bW_n)|=O_P(n^{-\nicefrac{1}{2}})$ as $n\to\infty$.
\end{lemma}
The proof of this lemma is provided in the supplementary material.
From this Lemma, we can investigate the properties of $\brho_n$ through its surrogate $\bW_n$. 
As $\bSigma_{\rho}$ represents the expectation of $\bW_n$, examining the structure of $\bSigma_{\rho}$ provides us with valuable insights of $\bW_n$. The spike structure of $\bSigma_{\rho}$ is illustrated in the following lemma. 
\begin{lemma}[Finite-rank perturbation]\label{lem:Sigma_finite_rank_perturbation}
	Under Assumptions \ref{assump-p-n}  --  \ref{assump-psi}, we have
	\begin{equation}\label{eq:Sigma_rho_decom-different-W}
		\Bigl\|\bSigma_{\rho}- \Bigl\{\diag\bigl(\bI_p-\gamma \bPsi^{-1}\bB\bB^{\top}\bPsi^{-1}\bigr) + \gamma \bPsi^{-1}\bB\bB^{\top}\bPsi^{-1}\Bigr\} \Bigr\|_2 = o(1)
	\end{equation}
	as $n\to \infty$, where 
    $\gamma := (6/\pi)\Expe\bigl[w^f/\{(w_1^e+w_2^e)(w_3^e+w_4^e)\}^{\nicefrac{1}{2}}\bigr]$, and $w_{j}^e$, $j=1,2,3,4$, are independent copies of $w_e$.
\end{lemma}

\begin{remark}
    Note that both $\bSigma_{\rho}$ and its approximation in \eqref{eq:Sigma_rho_decom-different-W} are correlation-type matrices, and all the diagonal entries equal to one. Consequently, the average of their eigenvalues are both one, and their bulk eigenvalues are clustered around one.
\end{remark}

The proof of this lemma is provided in the supplementary material.
In this lemma,  we derive a ``consistent'' approximation of the population covariance matrix $\bSigma_{\rho}$. From Weyl's lemma (Lemma S1.1 in the supplementary material),
the spectrum of the matrix $\bSigma_{\rho}$ can be approximated by that of a rank-$K$ perturbation of a diagonal matrix.
Intuitively, $\bSigma_{\rho}$ would have at most $K$ relatively larger eigenvalues. As for the sample counterpart, at most $K$ spiked sample eigenvalues of $\brho_n$ would lay outside the support of its LSD. 
Naturally, by counting the number of spiked eigenvalues of $\brho_n$, we can obtain a promising estimator of total number of factors. However, a very important yet intuitive observation here is that, for $j\in [K]$, $\lambda_j(\brho_n)$ is not always far away from the bulk eigenvalues $\{\lambda_j(\brho_n)\}_{j=K+1}^p$. It depends on whether the signal $\lambda_j(\bSigma_{\rho})$ is strong enough. If $\lambda_j(\bSigma_{\rho})$ is too weak, $\lambda_j(\brho_n)$ would lie on the boundary of the support of bulk eigenvalues.
This phenomenon is commonly referred to as the \emph{phase-transition phenomenon}, which is described in the following theorem.

\begin{theorem}[Phase transition]\label{thm:phase-transition}
     For the high-dimensional factor model \eqref{eq:factor-model}, assume that Assumptions \ref{assump-p-n}  --  \ref{assump-psi} hold, and the ESD of $\bSigma_{\rho}$ tends to a proper probability measure $H$ as $n\to\infty$. 
     Denoting $\psi(\alpha) = \alpha + c\int\frac{t\alpha}{\alpha-t}\dif H(t)$, we have
     \begin{description}
         \item[(a)] For $j\in [K]$ satisfying $\psi'\bigl(\lambda_j(\bSigma_{\rho})\bigr)>0$, the $j$-th sample eigenvalue of $\brho_n$ converges almost surely to $\psi\bigl(\lambda_j(\bSigma_{\rho})\bigr)$, which is outside the support of the LSD of $\brho_n$.
        \item[(b)]  For $j \in [K]$ satisfying $\psi'\bigl(\lambda_j(\bSigma_{\rho})\bigr)\leqslant 0$, 
       the $j$-th sample eigenvalue of $\brho_n$ converges almost surely to the right endpoint of the support of the LSD of $\brho_n$.  
     \end{description}
\end{theorem}
The proof of Theorem~\ref{thm:phase-transition} can be found in Section \ref{sec:lemma_proof_theorem}.

\begin{remark}\label{rmk:phase-transition-proof}
From Lemma \ref{lem:rho_W_spectral_norm}, we can investigate the asymptotic behavior of spiked eigenvalues of $\brho_n$ via those of $\bW_n$. 
Although $\bW_n=(3/n)\sum_{i=1}^n\bA_i\bA_i^{\top}$ is a Wishart-type random matrix, the nonlinear correlation structure of $\bA_i$ makes it difficult to directly apply the current phase-transition analysis techniques. 
The reason is that 
the vectors $\{\bA_i\}_{i=1}^n$ do not follow the commonly used \emph{independent component structure}  as in \citet{BY08,bai2012sample} and \citet{jiang2021generalized}. Specifically, the vector $\bA_i$ cannot be written as $\bA_i = \bSigma^{\nicefrac{1}{2}}\bx_i$, where $\bSigma$ is non-negative definite and all elements of $\bx_i\in\bbR^p$ are i.i.d. with zero mean and unit variance. 
To remove the independent component structure assumption, we first show that replacing $\{\sqrt{3}\bA_i\}_{i=1}^n$ in $\bW_n$ with i.i.d. $\calN_p(\boldsymbol{0},\bSigma_{\rho})$ random vectors does not change the asymptotic behavior of spiked eigenvalues of $\bW_n$ (see Lemma \ref{lem:G4MT} and Section \ref{sec:proof-phase-transition} for more details). 
The substitution of $\{\sqrt{3}\bA_i\}_{i=1}^n$ with i.i.d. $\calN_p(0,\bSigma_{\rho})$ is feasible because each entry of $\bA_i$ follows a  $\mathsf{Uniform}(-1,1)$ distribution and the universality phenomenon holds for light-tailed distributions. The universality phenomenon reveals that as long as $\bA_i$ has light-tailed entries, the first-order asymptotic behavior of the eigenvalues of $\frac{3}{n}\sum_{i=1}^n\bA_i\bA_i^{\top}$ remains the same when replacing $\{\sqrt{3}\bA_i\}_{i=1}^n$ with Gaussian vectors.
To guarantee the feasibility of this replacement, we establish concentration properties related to certain quadratic forms and their higher-order moments under the nonlinear correlation structure (see Lemma S1.7 in the supplementary material). 
Then since Gaussian random vectors naturally follows the independent component structure, we can directly apply the phase transition theory in \citet{BY08,bai2012sample} and \citet{jiang2021generalized} to complete the proof of Theorem \ref{thm:phase-transition}.
\end{remark}

To summarize, we first establish in Lemma \ref{lem:rho_W_spectral_norm} that $|\lambda_j(\rho_n)-\lambda_j(\bW_n)|=o_P(1)$, and subsequently turn to analyze the eigenvalues of $\bW_n$. Secondly, we prove that $\bSigma_{\rho}=\Expe \bW_n$ exhibits a rank-$K$ perturbation structure as in Lemma \ref{lem:Sigma_finite_rank_perturbation}. Thirdly, we confirm the phase-transition phenomenon for $\lambda_j(\bW_n)$, where $j\in [K]$. Therefore, the corresponding result of $\lambda_j(\brho_n)$ follows naturally, as demonstrated in Theorem \ref{thm:phase-transition}.

\subsection{Estimation of the number of factors}\label{sec:estimator}

With the phase-transition theory in Theorem \ref{thm:phase-transition}, we now propose our new estimator for the number of factors.
As stated in Theorem \ref{thm:phase-transition}, if $\psi'\bigl( \lambda_j(\bSigma_{\rho}) \bigr) \leqslant 0$ for some $j\in[K]$,  the corresponding sample eigenvalue $\lambda_j(\brho_n)$ will converge to the right endpoint of the support of the LSD of $\brho_n$, which is also the limit of the largest noise eigenvalue $\lambda_{K+1}(\brho_n)$. Hence, such weak factors will be merged into the noise component, making their signal undetectable.
By taking this into account, we define the \emph{number of significant factors} as
\begin{equation}\label{eq:significant-factor-number}
K_0 = \verb|#|\bigl\{j\in [K]: \psi'\bigl(\lambda_j(\bSigma_{\rho})\bigr) > 0 \bigr\},
\end{equation}
where the notation $\verb|#| \, \calS$ denotes the cardinality number of the set $\calS$. By Theorem \ref{thm:phase-transition}, the leading $K_0$  eigenvalues of $\brho_n$ will lay outside the support of its LSD . 

The LSD of $\brho_n$, denoted by $F_{c,H}$, is the generalized Mar\v{c}enko-Pastur law as stated in Lemma \ref{lem:rho_W_LSD}.
Let $\mathsf{supp}(F_{c,H})$ denote the support of $F_{c,H}$. 
The Stieltjes transform of $F_{c,H}$ is defined as $m(x) = \int \frac{1}{t-x} \dif F_{c,H}(t)$ for $x\in\bbR\setminus \mathsf{supp}(F_{c,H})$. Its first-order derivative $m'(x)$ is also only defined outside $\mathsf{supp}(F_{c,H})$, and can be extended as a function mapping the entire real line $\bbR$ to $\bbR\cup\{+\infty\}$ as follows:
\begin{equation}\label{eq:Stieltjes-derivative}
m'(x) = \begin{cases}
\int \frac{1}{(t-x)^2} \dif F_{c,H}(t),& \text{if }x\in\bbR\setminus \mathsf{supp}(F_{c,H}),\\
+\infty, & \text{if }x\in\mathsf{supp}(F_{c,H}).
\end{cases}
\end{equation}
This implies that $m'(\lambda_j(\brho_n))$ takes either finite or infinite values, depending on whether $\lambda_j(\brho_n)$ is a spiked eigenvalue or a bulk eigenvalue. Based on this observation, we utilize the derivative of the Stieltjes transform defined in \eqref{eq:Stieltjes-derivative} to identify all the spiked eigenvalues and estimate the total number of significant factors. Let $K_{\max}$ be a predetermined upper bound on the true number of significant factors, $K_0$. As the LSD $H$ of $\bSigma_{\rho}$ is unknown, we cannot obtain the explicit expression of $m'(x)$. Therefore, we utilize
\begin{equation}\label{eq:estimator-SR-Stieltje}
	\widehat{m}_{n,j}' (x) = \frac{1}{p-j} \sum_{\ell=j+1}^p\frac{1}{\bigl\{x - \lambda_{\ell}(\brho_n)\bigr\}^2}
\end{equation}
to estimate $m'(x)$ for $1\leqslant j \leqslant K_{\max}$. 
Intuitively, if $\lambda_{j}(\brho_n)$ lies within the bulk spectrum of $\brho_n$, we would expect $\widehat{m}_{n,j}' \bigl(\lambda_j(\brho_n)\bigr)$ to be very large. On the contrary, if $\lambda_{j}(\brho_n)$ is a spiked eigenvalue, $\widehat{m}_{n,j}' \bigl(\lambda_j(\brho_n)\bigr)$ should be relatively small. 
This phenomenon bears similarities to the behavior of $m'(x)$ described in the equation \eqref{eq:Stieltjes-derivative}. 
Actually, it will be shown that
\begin{equation}\label{eq:m_prime_hat}
\widehat{m}_{n,j}' \bigl( \lambda_j(\brho_n) \bigr)  
=\begin{cases}
    O_P(1), & \text{for } 1\leqslant j\leqslant K_0, \\
    O_P(p^{\nicefrac{1}{3}}), & \text{for } K_0+1\leqslant j\leqslant K_{\max},
\end{cases}
\end{equation}
as $n\to\infty$ (see the proof of Theorem \ref{thm:Stieltjes_ratio} in Section \ref{sec:proof-consistency}). 
Hence, a natural estimator of the number of significant factors is 
\begin{equation}\label{eq:estimator-SR}
\widehat{K}_{\SR} = \argmax_{1\leqslant j\leqslant K_{\max}}  \frac{\widehat{m}_{n,j+1}'(\lambda_{j+1}(\brho_n))}{\widehat{m}_{n,j}'(\lambda_{j}(\brho_n)) },
\end{equation}
where the ``S'' in subscript stands for Stieltjes transform, and the ``R'' stands for Ratio. 
\begin{remark}
	Here, we explain our preference for choosing $\widehat{m}_{n,j}'(\cdot)$ over directly employing eigen-ratio type estimators. 
	Under our current framework, both eigenvalues $\lambda_j(\brho_n)$ and the ratios $\frac{\lambda_{j+1}(\brho_n)}{\lambda_{j}(\brho_n)}$ are of constant order for all $1\leqslant j \leqslant K_{\max}$. The ratios of eigenvalues exhibit the following behavior:
	\[
		\frac{\lambda_{j+1}(\brho_n)}{\lambda_{j}(\brho_n)}\begin{cases}
			< 1, & 1\leqslant j \leqslant K_0,\\
			= 1 - \varepsilon_p, & K_0+1 \leqslant j \leqslant K_{\max},
		\end{cases}
	\]
	where $\varepsilon_p=o_P(1)$ and $\varepsilon_p>0$.  When factor signals are weak, the values of $\lambda_{j+1}(\brho_n)/\lambda_j(\brho_n)$ may be close to $1$ for both $j=K_0$ and $j=K_0+1$, making it difficult to distinguish between them. This can lead to inaccurate estimation.
    Performing the derivative of Stieltjes transform $\widehat{m}'_{n,j}(\cdot)$ on the corresponding eigenvalues $\lambda_j(\brho_n)$ significantly amplifies the ratio at $j=K_0$. As shown in Section 6.3, we found out that
	\begin{equation*}
		\frac{\widehat{m}_{n,j+1}'(\lambda_{j+1}(\brho_n))}{\widehat{m}_{n,j}'(\lambda_{j}(\brho_n))}
		\begin{cases}
			= O_P(1), & 1\leqslant j \leqslant K_0-1,\\
			\to \infty, & j= K_0,\\
			= O_P(1), & K_0+1 \leqslant j \leqslant K_{\max}.
		\end{cases}
	\end{equation*}
	As a result, the sequence of ratios $\{\widehat{m}_{n,j+1}'(\lambda_{j+1}(\brho_n))/\widehat{m}_{n,j}'(\lambda_{j}(\brho_n))\}$ blows up at $j=K_0$. 
	Therefore, estimating $K_0$ using $\widehat{m}_{n,j}'(\lambda_j(\brho_n))$ is more efficient compared to using ratios of $\lambda_j(\brho_n)$. Based on this observation, we propose the $\SR$ estimator.
\end{remark}

The consistency of this estimator is established in the following theorem, and its proof is postponed to Section \ref{sec:lemma_proof_theorem}.

\begin{theorem}[Consistency of $\widehat{K}_{\SR}$]\label{thm:Stieltjes_ratio}
    For the high-dimensional factor model \eqref{eq:factor-model}, assume that Assumptions \ref{assump-p-n}  --  \ref{assump-psi} hold. Let $K_0$ be the number of significant factors defined in \eqref{eq:significant-factor-number} and $\widehat{K}_{\SR}$ be the proposed estimator defined in \eqref{eq:estimator-SR-Stieltje} -- \eqref{eq:estimator-SR}. Then, we have 
    \[
    \lim_{n\to\infty}\Prob(\widehat{K}_{\SR}=K_0) = 1.
    \]
\end{theorem}

\section{Simulation studies}\label{sec:simu}

In this section, we conduct some simulations to examine the finite sample performance of the proposed estimator. 
We compare several estimators in the current literature, including our $\SR$ estimator; the $\NE$ estimator \citep{Nadakuditi2008sample}; the $\ED$ estimator \citep{onatski2010determining, Onatski2012asymptotic}; the $\BCV$ estimator \citep{Owen2016BCV}; the $\MKTCR$ estimator \citep{yu2019robust}, as well as the $\ACT$ estimator \citep{fan2020estimating}. The $\MKTCR$ estimator is designed to handle heavy-tailed data and can only detect strong factors. Conversely, other estimators are also capable of identifying weak factors. Specifically, these competing estimators are defined as follows:
\begin{enumerate}
	\item \underline{Our $\SR$ estimator:} 
	Here we use a slightly modified version of $\widehat{K}_{\SR}$ 	to circumvent numerical instability:
	\begin{equation}\label{eq:ktilde}
	\widetilde{K}_{\SR} = \argmax\limits_{1\leqslant j\leqslant K_{\max}}  \frac{\widetilde{m}_{n,j+1}'(\lambda_{j+1}(\brho_n))}{\widetilde{m}_{n,j}'(\lambda_{j}(\brho_n))}, 	
	\end{equation}
	where $\widetilde{m}_{n,j}'(\lambda_{j}(\brho_n)) = \frac{1}{p-j} \sum_{\ell=j+1}^p\frac{1}{\{\lambda_j(\brho_n) - \lambda_{\ell}(\brho_n)\}^2+p^{-\nicefrac{4}{3}}}$. Here we add  $p^{-\nicefrac{4}{3}}$ to the denominator in case that $\lambda_j(\brho_n) - \lambda_{\ell}(\brho_n)$ is too small. Similar to $\widehat{m}_{n,j}'(\lambda_{j}(\brho_n))$, $\widetilde{m}_{n,j}'(\lambda_{j}(\brho_n))$ still satisfies \eqref{eq:m_prime_hat}. $\widetilde{K}_{\SR}$ also retains the same consistency properties as $\widehat{K}_{\SR}$. Detailed proof is provided in Section \ref{sec:proof-consistency}.
%
	
	\item \underline{$\NE$ estimator:}
	Let $\{\by_i\}_{i=1}^n$ be an i.i.d. sample from the factor model \eqref{eq:factor-model}. The sample covariance matrix of $\{\by_i\}_{i=1}^n$ is defined as 
	$\bS_n  = n^{-1}\sum_{i=1}^n (\by_i-\bar{\by})(\by_i-\bar{\by})^{\top}$, where $\bar{\by}=n^{-1}\sum_{i=1}^n \by_i$ is the sample mean. Based on the eigenvalues of $\bS_n$,
    \citet{Nadakuditi2008sample} introduced the $\NE$ estimator as follows:
    \[
        \widehat{K}_{\NE} = \argmin_{0\leqslant j < \min(p,n)} \biggl\{ \frac{1}{4}\biggl(\frac{n}{p}\biggr)^2 t_j^2 + 2(j+1) \biggr\},
    \]  
    where 
    $
        t_j = p\bigl[(p-j) \{\sum_{i=j+1}^p \lambda_i(\bS_n) \}^{-2}\sum_{i=j+1}^p \lambda_i^2(\bS_n) -1 - p/n \bigr] - p/n.
    $
    
	\item \underline{$\ED$ estimator:}
	Based on the eigenvalues of $\bS_n$, \citet{onatski2010determining, Onatski2012asymptotic} proposed an eigenvalue difference criterion, defined as
    \[
        \widehat{K}_{\ED} = \max \{1\leqslant j \leqslant K_{\max}, \lambda_j(\bS_n)-\lambda_{j+1}(\bS_n) \geqslant \delta \},
    \]
    where $\delta$ is a predetermined threshold calculated using a calibration method described in \citet[Section IV]{onatski2010determining}. 
	
	\item \underline{$\BCV$ estimator:}
	\citet{Owen2016BCV} introduced an algorithm to determine the number of factors based on $\bS_n$ and the bi-cross-validation (BCV) technique from \citet{Owen2009BCV}. This method involves randomly holding out some rows and some columns of the observed data, fitting a factor model to the held-in data, and comparing held-out data to corresponding fitted values. We utilize Owen and Wang's R package ``\textsf{esaBcv}'' to implement the $\BCV$ method in our simulation studies.

	\item \underline{$\MKTCR$ estimators:}
  The sample multivariate Kendall's tau matrix is defined as
	$
	\bK_n = \frac{2}{n(n-1)} \sum_{1\leqslant i<\ell\leqslant n} \frac{(\by_{i}-\by_{\ell})(\by_{i}-\by_{\ell})^{\top}}{\|\by_{i}-\by_{\ell}\|^2}.
	$
	Based on the eigenvalues of $\bK_n$, \cite{yu2019robust} constructed the $\MKTCR$ estimator as follows:
	\[
            \widehat{K}_{\MKTCR} = \argmax_{1\leqslant j\leqslant K_{\max}} \frac{\ln\{ 1+\lambda_j(\bK_n)/V_{j-1} \}}{\ln\{ 1+\lambda_{j+1}(\bK_n)/V_{j} \}},
	\]
	where
	$V_j = \sum_{i=j+1}^{\min(p,n)} \lambda_i(\bK_n)$, $0\leqslant j \leqslant \min(p,n)-1$.
	
	\item \underline{$\ACT$ estimator:}
	The sample Pearson correlation matrix of $\{\by_i\}_{i=1}^n$ is defined as
    $
		\bP_n  = \bigl[\diag(\bS_n)\bigr]^{-\nicefrac{1}{2}}\bS_n\bigl[\diag(\bS_n)\bigr]^{-\nicefrac{1}{2}}.
    $
	Based on the spectral properties of $\bP_n$, \cite{fan2020estimating} proposed an estimator to estimate the factor number as follows: 
	\[
	\widehat{K}_{\ACT} = \max\Bigl\{ 1\leqslant j\leqslant K_{\max}: \widehat{\alpha}_j(\bP_n) > 1+\sqrt{p/(n-1)} \Bigr\},
	\]
	where $\{\widehat{\alpha}_j(\bP_n)\}_{j=1}^{p}$ are bias correction of sample eigenvalues of $\bP_n$, defined as
	$
	\widehat{\alpha}_j(\bP_n) = -1/\underline{m}_{n,j}(\lambda_j(\bP_n))
	$
	with $\underline{m}_{n,j}(x)  = -(1-c_j)/x + c_j m_{n,j}(x)$, $c_j = (p-j)/(n-1)$, and
	\[
		m_{n,j}(x)  = \frac{1}{p-j} \Biggl[ \sum_{\ell=j+1}^p \frac{1}{\lambda_{\ell}(\bP_n)-x} + \frac{1}{\{3\lambda_{j}(\bP_n)+\lambda_{j+1}(\bP_n)\}/4-x}\Biggr].
	\]
\end{enumerate}

Our simulation studies consider various combinations of dimension and sample size, namely 
$(p,n) = (50,100),$ $ (100,200)$, $(150,300)$, and $ (200,400)$, which all have the ratio $p/n=1/2$. We take the true number of common factors $K=3$ and set the possible maximum value of the number of common factors $K_{\max}=20$. 
Recalling our distribution assumption \ref{assump-SMN-different-W} for both common factors $\bff_i$ and idiosyncratic errors $\be_i$, we generate $\bff_i$ and $\be_i$ by $\bff_i = (w_i^f)^{\nicefrac{1}{2}} \bx_i$ and $e_{ij} = (w_{ij}^e)^{\nicefrac{1}{2}}z_{ij},$
where $e_{ij}$ denotes the $j$-th component of $\be_i$, $\{\bx_i\}_{i=1}^n\iidsim \calN_K(\boldsymbol{0},\bI_K)$, and $\{z_{ij}, i\in[n], j\in[p]\}\iidsim \calN(0,1)$. We employ four different scenarios to generate sample data for $w_i^f$ and $w_{ij}^e$:
\begin{enumerate}
    \item (Normal population, see Table \ref{tab:simu_normal_mix}) Let $w_i^f=w_{ij}^e=1$ for all $i\in[n]$ and $j\in[p]$;
    \item (Uniform and Chi-squared population, see Table \ref{tab:simu_normal_mix}) Let $\{w_i^f\}_{i=1}^n\iidsim\mathsf{Uniform}(0,1)$ and $\{w_{ij}^e, i\in[n], j\in[p]\}\iidsim \chi^2(1)$;
    \item (Student's $t(2)$ population, see Table \ref{tab:simu_t2_cauchy}) Let $\{w_i^f\}_{i=1}^n\iidsim\invGamma(1,1)$ and $\{w_{ij}^e, i\in[n], j\in[p]\}\iidsim \invGamma(1,1)$, where $\invGamma(\alpha, \beta)$ denotes the inverse Gamma distribution with shape parameter $\alpha$ and scale parameter $\beta$. In this scenario, both $\bff_i$ and $e_{ij}$ follow (multivariate) Student's $t(2)$ distributions;
    \item (Cauchy population, see Table \ref{tab:simu_t2_cauchy}) Let $\{w_i^f\}_{i=1}^n\iidsim\invGamma(1/2,1/2)$ and $\{w_{ij}^e, i\in[n], j\in[p]\}\iidsim \invGamma(1/2,1/2)$. In this scenario, both $\bff_i$ and $e_{ij}$ follow (multivariate) Cauchy distribution.
\end{enumerate}

Furthermore, we consider three cases for the loading matrix $\bB=(B_{ij})_{p\times K}$ and the matrix $\bPsi$ as follows. (C1) is from \citep{Harding2013estimating, fan2020estimating}. (C2) and (C3) are both from \citet{Onatski2012asymptotic}.
\begin{enumerate}[label={(C\arabic*)}, ref={(Case~\arabic*)}]
    \item For any $j\in[K]$, let $B_{ij}=\sqrt{5j/p}$ for $i\in[K]$, and let $B_{ij}=a_{ij}\sqrt{5j/(p-j)}$ for $i\in\{K+1,\ldots,p\}$, where $a_{ij}=-1$ if $i=rj$ or $a_{ij}=1$ if $i\neq r j$, $r\in\bbN^+$. Let $\bPsi = \bI_p$. 

    \item For any $i\in [p]$ and $j\in [K]$, let $\sqrt{p}B_{ij}/\sqrt{10 j}\iidsim \calN(0,1)$. Let $\bPsi = \bI_p$. 
		
		\item For any $i\in [p]$ and $j\in [K]$, let $\sqrt{p}B_{ij}/\sqrt{10 j}\iidsim \calN(0,1)$. Let $\bPsi = \bT^{\nicefrac{1}{2}}$, where $\bT$ is a Toeplitz matrix with its $(i,j)$-th entry equal to $0.45^{|i-j|}$.  
\end{enumerate}

\begin{figure}[]
    \centering
    \includegraphics[width=\textwidth]{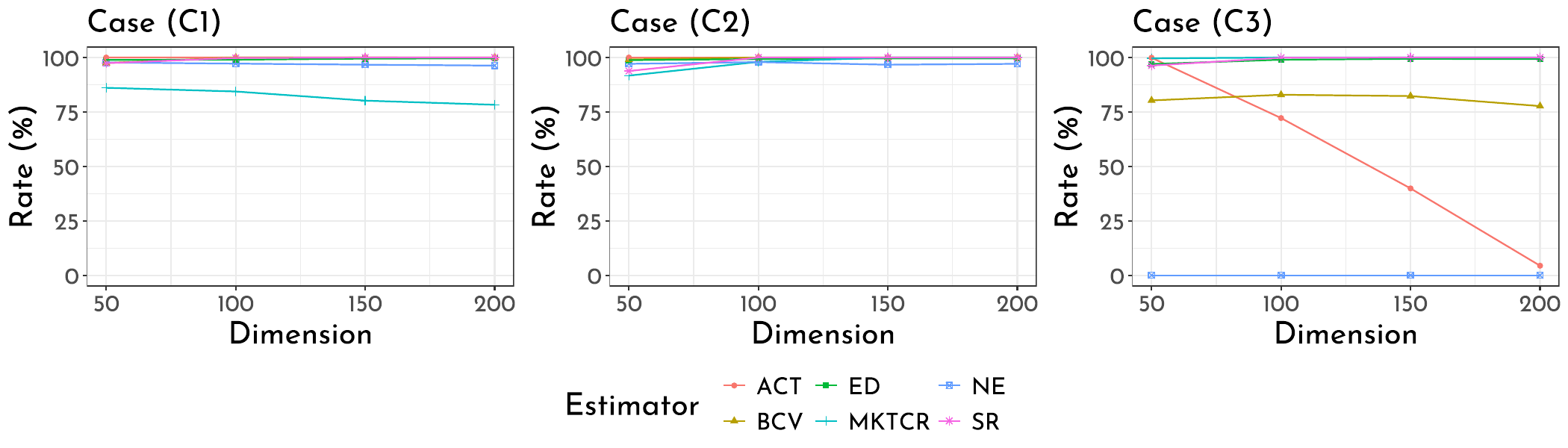}
    \caption{Correct identification rate of six estimators. Entries of common factors and idiosyncratic errors are generated from the standard normal distribution.}
    \label{fig:rate-normal}
\end{figure}

\begin{figure}[]
    \centering
    \includegraphics[width=\textwidth]{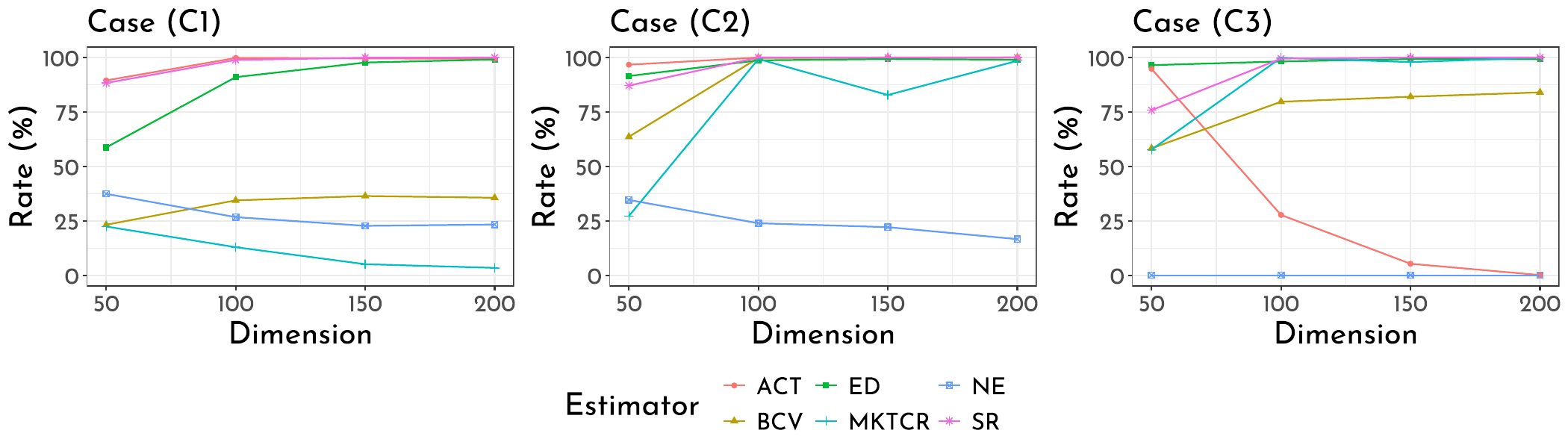}
    \caption{Correct identification rate of six estimators. Entries of common factors are generated from a scale mixture of normals with $\{w_i^f\}_{i=1}^n\iidsim \mathsf{Uniform}(0,1)$, and entries of idiosyncratic errors are generated from a scale mixture of normals with $\{w_{ij}^e,i\in[n],j\in[p]\}\iidsim \chi^2(1)$.}
    \label{fig:rate-mix}
\end{figure}

\begin{figure}[]
    \centering
    \includegraphics[width=\textwidth]{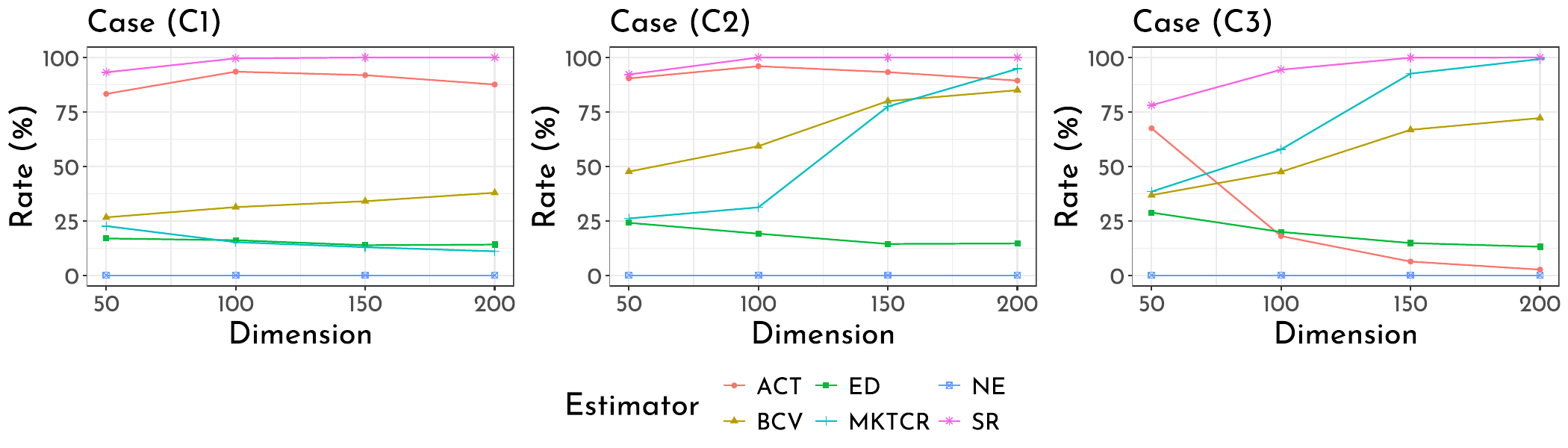}
    \caption{Correct identification rate of six estimators. Entries of common factors and idiosyncratic errors are generated from Student's $t(2)$ distribution.}
    \label{fig:rate-t2}
\end{figure}

\begin{figure}[]
    \centering
    \includegraphics[width=\textwidth]{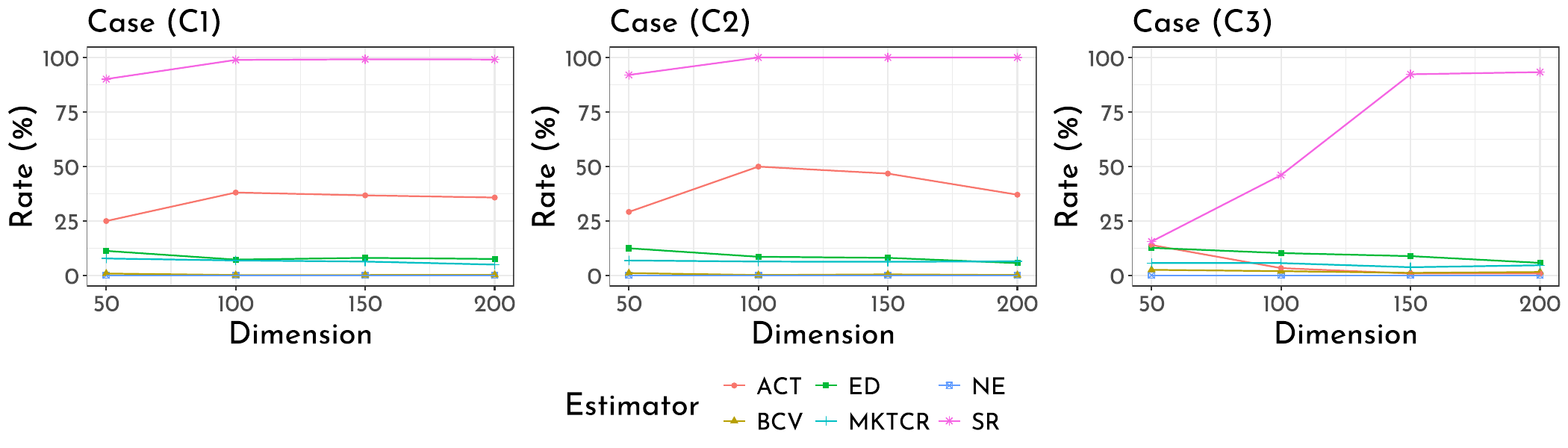}
    \caption{Correct identification rate of six estimators. Entries of common factors and idiosyncratic errors are generated from the standard Cauchy distribution.}
    \label{fig:rate-cauchy}
\end{figure}

The simulation results are reported in Tables \ref{tab:simu_normal_mix} - \ref{tab:simu_t2_cauchy} and Figures \ref{fig:rate-normal} - \ref{fig:rate-cauchy}. 
In the case of light-tailed data (see Tables \ref{tab:simu_normal_mix} and Figures \ref{fig:rate-normal} - \ref{fig:rate-mix}), our $\SR$ estimator performs
comparably to other estimators. However, when handling heavy-tailed data (see Tables \ref{tab:simu_t2_cauchy} and Figures \ref{fig:rate-t2} - \ref{fig:rate-cauchy}), the $\NE$, $\ED$, and $\BCV$ estimators, which are based on sample covariance matrices, prove ineffective, whereas our
$\SR$ estimator outperforms the $\MKTCR$ and $\ACT$ estimators.

\begin{table}[htbp]
  \scriptsize
	\centering 
	\caption{Percentages (\%) of estimated number of common factors in $1000$ simulations. Entries of common factors and idiosyncratic errors are generated from \textbf{light-tailed}  distributions. The results are reported in the form $a(b\vert c)$, in which $a, b, c$ are the percentages of true estimates, overestimates, and underestimates, respectively. The notation ``$\mathrm{ave}(\widehat{K})$'' denotes mean estimators for the case $(p,n)=(200,400)$.}
    \begin{tabular}{ccllllll}
    \toprule
    \textbf{Case} & $p$   & \textbf{\texttt{NE}} & \textbf{\texttt{ED}} & \textbf{\texttt{BCV}} & \textbf{\texttt{MKTCR}} & \textbf{\texttt{ACT}} & \textbf{\texttt{SR}} \\
    \midrule
          & \multicolumn{7}{c}{\textbf{Normal population}} \\
\cmidrule{2-8}    \multirow{5}[4]{*}{(C1)} & 50    & 97.7(2.3$|$0) & 98.9(1.1$|$0) & 97.5(0.5$|$2) & 86.1(0$|$13.9) & 100(0$|$0) & \textbf{97.4}(0.5$|$2.1) \\
          & 100   & 97.2(2.8$|$0) & 99(1$|$0) & 100(0$|$0) & 84.4(0$|$15.6) & 100(0$|$0) & \textbf{99.8}(0$|$0.2) \\
          & 150   & 96.7(3.3$|$0) & 99.5(0.5$|$0) & 100(0$|$0) & 80.2(0$|$19.8) & 100(0$|$0) & \textbf{100}(0$|$0) \\
          & 200   & 96.3(3.7$|$0) & 99.6(0.4$|$0) & 100(0$|$0) & 78.3(0$|$21.7) & 99.9(0.1$|$0) & \textbf{100}(0$|$0) \\
\cmidrule{2-8}          & $\mathrm{ave}(\widehat{K})$ & 3.038 & 3.004 & 3     & 2.783 & 3.001 & \textbf{3} \\
    \midrule
    \multirow{5}[4]{*}{(C2)} & 50    & 97.1(2.9$|$0) & 98.8(1.2$|$0) & 99.1(0.8$|$0.1) & 91.7(0$|$8.3) & 100(0$|$0) &\textbf{93.8}(0$|$6.2) \\
          & 100   & 97.9(2.1$|$0) & 99.3(0.7$|$0) & 100(0$|$0) & 98(0$|$2) & 100(0$|$0) & \textbf{100}(0$|$0) \\
          & 150   & 96.7(3.3$|$0) & 99.6(0.4$|$0) & 100(0$|$0) & 100(0$|$0) & 100(0$|$0) &  \textbf{100}(0$|$0) \\
          & 200   & 97.1(2.9$|$0) & 99.6(0.4$|$0) & 100(0$|$0) & 100(0$|$0) & 100(0$|$0) & \textbf{100}(0$|$0) \\
\cmidrule{2-8}          & $\mathrm{ave}(\widehat{K})$ & 3.03  & 3.007 & 3     & 3     & 3     & \textbf{3} \\
    \midrule
    \multirow{5}[4]{*}{(C3)} & 50    & 0(100$|$0) & 97(3$|$0) & 80.3(19.7$|$0) & 99.6(0$|$0.4) & 99.9(0.1$|$0) & \textbf{96.3}(0.3$|$3.4) \\
          & 100   & 0(100$|$0) & 99(1$|$0) & 82.9(17.1$|$0) & 100(0$|$0) & 72.2(27.8$|$0) & \textbf{99.9}(0.1$|$0) \\
          & 150   & 0(100$|$0) & 99.4(0.6$|$0) & 82.3(17.7$|$0) & 100(0$|$0) & 39.9(60.1$|$0) & \textbf{100}(0$|$0) \\
          & 200   & 0(100$|$0) & 99.3(0.7$|$0) & 77.7(22.3$|$0) & 100(0$|$0) & 4.4(95.6$|$0) & \textbf{100}(0$|$0) \\
\cmidrule{2-8}          & $\mathrm{ave}(\widehat{K})$ & 55.225 & 3.008 & 3.28  & 3     & 5.448 & \textbf{3} \\
    \midrule
          & \multicolumn{7}{c}{\textbf{Uniform and Chi-squared population}} \\
\cmidrule{2-8}    \multirow{5}[4]{*}{(C1)} & 50    & 37.4(62.3$|$0.3) & 58.7(2.7$|$38.6) & 23.2(0.4$|$76.4) & 22.4(0$|$77.6) & 89.5(0$|$10.5) & \textbf{88.3}(0.6$|$11.1) \\
          & 100   & 26.7(73.3$|$0) & 91(1.1$|$7.9) & 34.4(0$|$65.6) & 12.9(0$|$87.1) & 99.8(0.1$|$0.1) & \textbf{98.9}(0.1$|$1) \\
          & 150   & 22.7(77.2$|$0.1) & 97.7(0.3$|$2) & 36.4(0.1$|$63.5) & 5.1(0$|$94.9) & 99.6(0.4$|$0) & \textbf{99.9}(0$|$0.1) \\
          & 200   & 23.3(76.7$|$0) & 99.1(0.8$|$0.1) & 35.6(0$|$64.4) & 3.4(0$|$96.6) & 99.4(0.6$|$0) & \textbf{100}(0$|$0) \\
\cmidrule{2-8}          & $\mathrm{ave}(\widehat{K})$ & 4.204 & 3.009 & 2.356 & 1.919 & 3.006 & \textbf{3} \\
    \midrule
    \multirow{5}[4]{*}{(C2)} & 50    & 34.6(65.3$|$0.1) & 91.5(2.4$|$6.1) & 63.6(1.7$|$34.7) & 27.2(0$|$72.8) & 96.7(0$|$3.3) & \textbf{87.1}(0.2$|$12.7) \\
          & 100   & 23.9(76.1$|$0) & 98.7(1.3$|$0) & 99.8(0.2$|$0) & 99.3(0$|$0.7) & 100(0$|$0) & \textbf{100}(0$|$0) \\
          & 150   & 22.1(77.9$|$0) & 99.3(0.7$|$0) & 100(0$|$0) & 82.8(0$|$17.2) & 99.8(0.2$|$0) & \textbf{100}(0$|$0) \\
          & 200   & 16.6(83.4$|$0) & 99(1$|$0) & 100(0$|$0) & 98.5(0$|$1.5) & 100(0$|$0) & \textbf{100}(0$|$0) \\
\cmidrule{2-8}          & $\mathrm{ave}(\widehat{K})$ & 4.343 & 3.01  & 3     & 2.985 & 3     & \textbf{3} \\
    \midrule
    \multirow{5}[4]{*}{(C3)} & 50    & 0(100$|$0) & 96.5(2.6$|$0.9) & 58.3(41.6$|$0.1) & 57.7(0$|$42.3) & 94.8(5.2$|$0) & \textbf{75.8}(2.2$|$22) \\
          & 100   & 0(100$|$0) & 98.2(1.8$|$0) & 79.7(20.3$|$0) & 99.9(0$|$0.1) & 27.7(72.3$|$0) & \textbf{99.5}(0.5$|$0) \\
          & 150   & 0(100$|$0) & 99.3(0.7$|$0) & 82(18$|$0) & 97.9(0$|$2.1) & 5.3(94.7$|$0) & \textbf{100}(0$|$0) \\
          & 200   & 0(100$|$0) & 99.2(0.8$|$0) & 84(16$|$0) & 100(0$|$0) & 0.1(99.9$|$0) & \textbf{100}(0$|$0) \\
\cmidrule{2-8}          & $\mathrm{ave}(\widehat{K})$ & 56.198 & 3.01  & 3.193 & 3     & 7.554 & \textbf{3} \\
    \bottomrule
    \end{tabular}
	\label{tab:simu_normal_mix}%
\end{table}%

\begin{table}[htbp]
    \scriptsize
    \centering
    \caption{Percentages (\%) of estimated number of common factors in $1000$ simulations. Entries of common factors and idiosyncratic errors are generated from \textbf{heavy-tailed} distributions. The results are reported in the form $a(b\vert c)$, in which $a, b, c$ are the percentages of true estimates, overestimates, and underestimates, respectively. The notation ``$\mathrm{ave}(\widehat{K})$'' denotes mean estimators for the case $(p,n)=(200,400)$.}
    \begin{tabular}{ccllllll}
    \toprule
    \textbf{Case} & $p$   & \textbf{\texttt{NE}} & \textbf{\texttt{ED}} & \textbf{\texttt{BCV}} & \textbf{\texttt{MKTCR}} & \textbf{\texttt{ACT}} & \textbf{\texttt{SR}} \\
    \midrule
          & \multicolumn{7}{c}{\textbf{$t(2)$ population}} \\
\cmidrule{2-8}    \multirow{5}[4]{*}{(C1)} & 50    & 0(100$|$0) & 16.9(29.4$|$53.7) & 26.6(6.3$|$67.1) & 22.6(0.2$|$77.2) & 83.3(0.3$|$16.4) & \textbf{93.2}(0$|$6.8) \\
          & 100   & 0(100$|$0) & 16.1(33.6$|$50.3) & 31.3(4$|$64.7) & 15.2(0$|$84.8) & 93.5(4.3$|$2.2) & \textbf{99.6}(0$|$0.4) \\
          & 150   & 0(100$|$0) & 13.8(33.2$|$53) & 34(4.2$|$61.8) & 12.9(0$|$87.1) & 91.9(7.4$|$0.7) & \textbf{100}(0$|$0) \\
          & 200   & 0(100$|$0) & 14.1(38$|$47.9) & 37.9(2.9$|$59.2) & 11(0$|$89) & 87.6(11.6$|$0.8) & \textbf{100}(0$|$0) \\
\cmidrule{2-8}          & $\mathrm{ave}(\widehat{K})$ & 37.884 & 3.316 & 2.228 & 1.941 & 3.12  & \textbf{3} \\
    \midrule
    \multirow{5}[4]{*}{(C2)} & 50    & 0(100$|$0) & 24.1(44.2$|$31.7) & 47.6(10.6$|$41.8) & 26.1(0.2$|$73.7) & 90.4(0.3$|$9.3) & \textbf{92.2}(0$|$7.8) \\
          & 100   & 0(100$|$0) & 19.1(44.9$|$36) & 59.3(8.3$|$32.4) & 31.2(0$|$68.8) & 96(2.9$|$1.1) & \textbf{100}(0$|$0) \\
          & 150   & 0(100$|$0) & 14.4(49.7$|$35.9) & 80(4$|$16) & 77.5(0$|$22.5) & 93.3(6.6$|$0.1) & \textbf{100}(0$|$0) \\
          & 200   & 0(100$|$0) & 14.6(53.1$|$32.3) & 85(2.1$|$12.9) & 94.8(0$|$5.2) & 89.4(10.5$|$0.1) & \textbf{100}(0$|$0) \\
\cmidrule{2-8}          & $\mathrm{ave}(\widehat{K})$ & 38.064 & 4.501 & 2.836 & 2.944 & 3.109 & \textbf{3} \\
    \midrule
    \multirow{5}[4]{*}{(C3)} & 50    & 0(100$|$0) & 28.8(50.1$|$21.1) & 36.8(49.5$|$13.7) & 38.4(0.1$|$61.5) & 67.5(28.8$|$3.7) & \textbf{78.1}(2.3$|$19.6) \\
          & 100   & 0(100$|$0) & 19.9(57.1$|$23) & 47.5(43.6$|$8.9) & 57.8(0$|$42.2) & 18(81.4$|$0.6) & \textbf{94.4}(2.8$|$2.8) \\
          & 150   & 0(100$|$0) & 14.8(64.5$|$20.7) & 66.8(28.6$|$4.6) & 92.6(0$|$7.4) & 6.3(93.5$|$0.2) & \textbf{99.9}(0.1$|$0) \\
          & 200   & 0(100$|$0) & 13.1(66.3$|$20.6) & 72.2(24.8$|$3) & 99.3(0$|$0.7) & 2.6(97.3$|$0.1) & \textbf{100}(0$|$0) \\
\cmidrule{2-8}          & $\mathrm{ave}(\widehat{K})$ & 70.59 & 4.911 & 3.47  & 2.993 & 7.959 & \textbf{3} \\
    \midrule
          & \multicolumn{7}{c}{\textbf{Cauchy population}} \\
\cmidrule{2-8}    \multirow{5}[4]{*}{(C1)} & 50    & 0(100$|$0) & 11.2(77.6$|$11.2) & 0.8(0.3$|$98.9) & 7.7(5.9$|$86.4) & 24.9(2.1$|$73) & \textbf{90.1}(0.2$|$9.7) \\
          & 100   & 0(100$|$0) & 7.2(85$|$7.8) & 0.1(0.1$|$99.8) & 6.8(5$|$88.2) & 38(12.4$|$49.6) & \textbf{98.9}(0.2$|$0.9) \\
          & 150   & 0(100$|$0) & 8(84.4$|$7.6) & 0.1(0$|$99.9) & 6.3(4.2$|$89.5) & 36.7(28.4$|$34.9) & \textbf{99.2}(0.7$|$0.1) \\
          & 200   & 0(100$|$0) & 7.5(86.5$|$6) & 0.2(0.1$|$99.7) & 4.9(4.8$|$90.3) & 35.7(41.7$|$22.6) & \textbf{99.1}(0.9$|$0) \\
\cmidrule{2-8}          & $\mathrm{ave}(\widehat{K})$ & 118.068 & 8.138 & 0.038 & 1.507 & 3.439 & \textbf{3.009} \\
    \midrule
    \multirow{5}[4]{*}{(C2)} & 50    & 0(100$|$0) & 12.4(76.9$|$10.7) & 1(0.7$|$98.3) & 6.8(2.6$|$90.6) & 29.1(2.3$|$68.6) & \textbf{92}(0$|$8) \\
          & 100   & 0(100$|$0) & 8.5(83.3$|$8.2) & 0.1(0.1$|$99.8) & 6.3(3.2$|$90.5) & 49.9(17$|$33.1) & \textbf{100}(0$|$0) \\
          & 150   & 0(100$|$0) & 8(86.8$|$5.2) & 0.4(0$|$99.6) & 6.2(2.1$|$91.7) & 46.7(35.6$|$17.7) & \textbf{100}(0$|$0) \\
          & 200   & 0(100$|$0) & 5.6(88$|$6.4) & 0.1(0.2$|$99.7) & 6.4(2.8$|$90.8) & 37(49.5$|$13.5) & \textbf{100}(0$|$0) \\
\cmidrule{2-8}          & $\mathrm{ave}(\widehat{K})$ & 118.511 & 8.236 & 0.069 & 1.486 & 3.596 & \textbf{3} \\
    \midrule
    \multirow{5}[4]{*}{(C3)} & 50    & 0(100$|$0) & 12.6(77.5$|$9.9) & 2.5(10.1$|$87.4) & 5.6(3.8$|$90.6) & 13.9(74.6$|$11.5) & \textbf{15.5}(17.8$|$66.7) \\
          & 100   & 0(100$|$0) & 10.2(81$|$8.8) & 1.9(4.7$|$93.4) & 5.6(1.4$|$93) & 3.3(93.4$|$3.3) & \textbf{46}(27.3$|$26.7) \\
          & 150   & 0(100$|$0) & 8.8(83.5$|$7.7) & 1.1(4.8$|$94.1) & 3.7(1.3$|$95) & 0.8(97.3$|$1.9) & \textbf{92.3}(7.5$|$0.2) \\
          & 200   & 0(100$|$0) & 5.7(87.2$|$7.1) & 1.5(3.5$|$95) & 4.6(1.2$|$94.2) & 0.8(98.5$|$0.7) & \textbf{93.3}(6.5$|$0.2) \\
\cmidrule{2-8}          & $\mathrm{ave}(\widehat{K})$ & 126.004 & 7.919 & 0.421 & 1.416 & 16.854 & \textbf{3.268} \\
    \bottomrule
    \end{tabular}
  \label{tab:simu_t2_cauchy}
\end{table}%

\section{Real data analysis}\label{sec:real-data-example}

In this section, we analyze the monthly macroeconomic dataset (FRED-MD, \cite{mccracken2016fred}) from March 1959 to January 2023. 
The data can be downloaded from the website \url{http://research.stlouisfed.org/econ/mccracken/fred-md/}, and includes the monthly series of 128 macroeconomic variables.  Following \cite{mccracken2016fred}, the series with missing values are removed and the remaining dataset is transformed to a stationary form. After this preprocessing procedure,  the data dimension is $p = 105$ and the sample size is $n = 767$. \cite{mccracken2016fred}'s recommendation to remove outliers has not been implemented in our data analysis, as we believe that data with heavy-tailed distributions will inevitably contain extreme observations that cannot be circumvented. Since our estimator is tailored to heavy-tailed observations, we directly use it to identify the number of factors.

\begin{figure}[htbp]
	\centering
	\begin{subfigure}{.45\textwidth}
		\includegraphics[width=\textwidth]{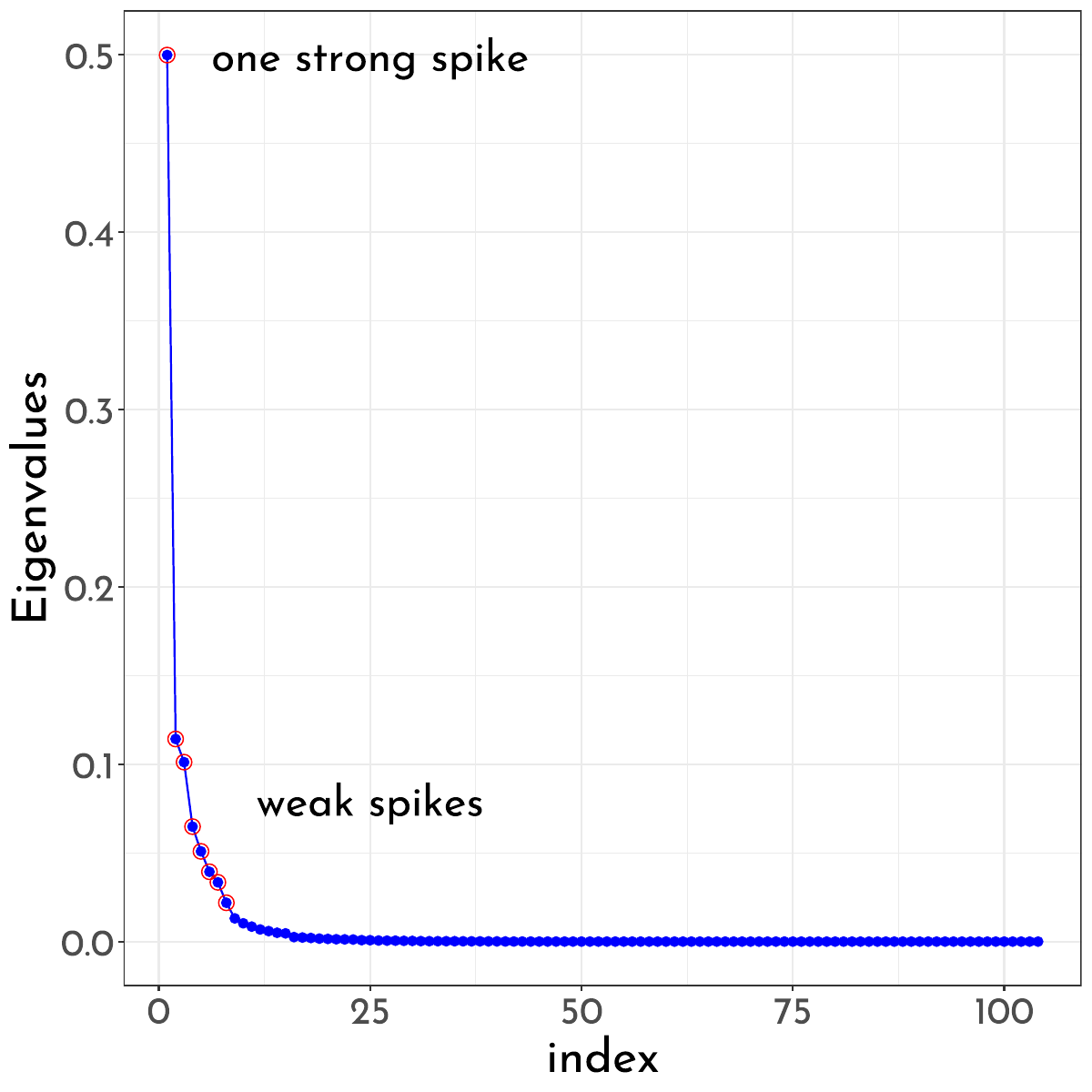}
		\caption{Multivariate Kendall's tau matrix}
		\label{fig:RealData-MKen}
	\end{subfigure}%
        \qquad
	\begin{subfigure}{.45\textwidth}
		\includegraphics[width=\textwidth]{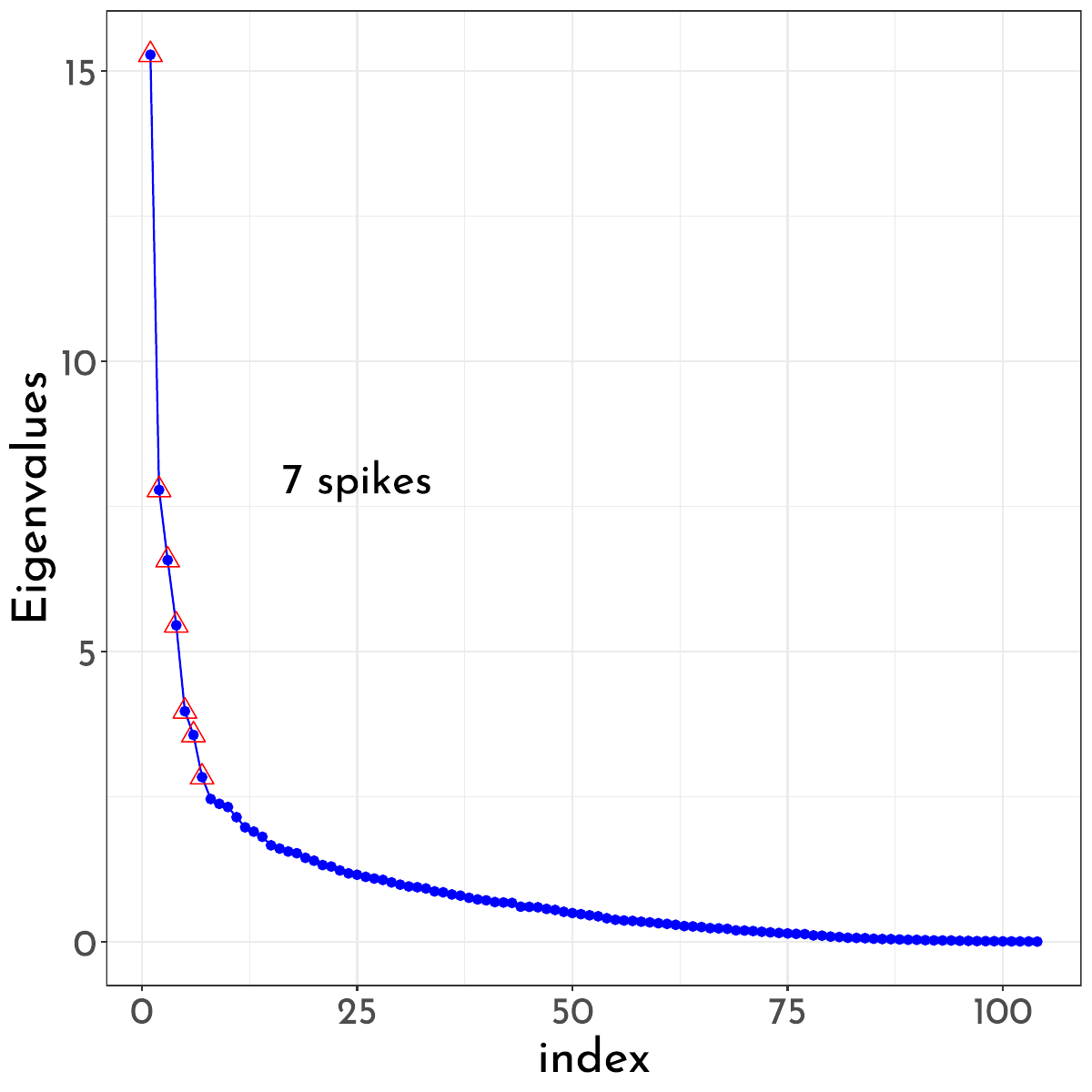}
		\caption{Spearman correlation matrix}
		\label{fig:RealData-Spearman}
	\end{subfigure}%

	\caption{Scatter plots of all the eigenvalues of multivariate Kendall's tau matrix and Spearman correlation matrix generated from the real dataset. The $\MKTCR$ estimator only detects one ``strong'' spiked eigenvalue of the multivariate Kendall's tau matrix, and neglects several ``weak'' spikes. Our $\SR$ estimator detects all seven spikes of Spearman correlation matrix.}
    \label{fig:RealData-eigvals-scatter}
\end{figure}

The dataset reveals that more than $67\%$ of the macroeconomic variables exhibit a sample kurtosis that exceeds 9, which is the theoretical kurtosis of the Student's $t(5)$ distribution. This indicates that the dataset is probably heavy-tailed. Compared to other estimators, $\MKTCR$, $\ACT$, and $\SR$ have slightly higher accuracy under heavy-tailed conditions, so we employ these three methods for estimation. The results are as follows: $\widehat{K}_{\ACT}=13$, $\widehat{K}_{\MKTCR}=1$, and $\widehat{K}_{\SR}=7$. 
As shown in the simulation studies in Section \ref{sec:simu}, $\ACT$ has similar performance to our estimator when data is light-tailed, while it tends to overestimate when data is heavy-tailed. The same story happens for this real dataset. Both our $\SR$ estimator and \citet{yu2019robust}'s $\MKTCR$ estimator are based on eigenvalues of certain type of sample correlation  matrices as plotted in Figure \ref{fig:RealData-eigvals-scatter}. From Figure \ref{fig:RealData-eigvals-scatter}\subref{fig:RealData-MKen}, it is evident that the multivariate Kendall's tau matrix exhibits one ``strong'' spike and several ``weak'' spikes. However, the $\MKTCR$ estimator only detects the strong spike while ignoring the weaker spikes. It potentially underestimates the total number of factors, similarly as shown in the simulation studies in Section \ref{sec:simu}. 
On the other hand, Figure \ref{fig:RealData-eigvals-scatter}\subref{fig:RealData-Spearman} illustrates that our $\SR$ estimator has successfully detected all seven spikes of the Spearman sample correlation matrix.
Therefore,  $\widehat{K}_{\SR}=7$ is a more persuasive estimation for this dataset.

\section{Discussions}\label{sec:discussion}

In summary, we propose a novel estimator to identify the number of common factors in high-dimensional factor models when the data is heavy-tailed. We demonstrate that, under certain assumptions, the number of spiked eigenvalues of the Spearman sample correlation matrix is consistent with the total number of significant factors. Our estimator is constructed based on this observation, and its consistency is proved under mild assumptions. From the perspective of RMT, we investigate the eigenstructure of the Spearman sample correlation matrix under spike models and establish the phase-transition theory of its spiked eigenvalues. Simulation results demonstrate that our proposed estimator outperforms competing methods in various scenarios, especially with heavy-tailed observations.
However, our $\SR$ estimator does not perform well when the sample size is not large enough, such as when $(p,n)=(50,100)$. The possible reason is that the estimation of $m'(x)$ is inaccurate when the sample size is small. A more accurate estimator for $m'(x)$ would improve the accuracy of our $\SR$ estimator. Furthermore, it is worth extending our method for factor modeling in high-dimensional time series \citep{lam2012factor,li2017identifying} and tensor data \citep{lam2021rank, chen2024rank}.
These extensions are beyond the scope of the current paper, and we leave them for future work.

\section{Proof of Theorems \ref{thm:phase-transition} and \ref{thm:Stieltjes_ratio} }\label{sec:lemma_proof_theorem}

\subsection{Some technical lemmas}\label{sec:lemmas}

In this section, we propose two technical lemmas in preparation for proving Theorems \ref{thm:phase-transition} and \ref{thm:Stieltjes_ratio}. The proofs of these lemmas are relegated to supplementary material.

Lemma \ref{lem:rho_W_LSD} provides the LSD of $\brho_n$ and $\bW_n$, extending the  result of \cite{wu2022limiting}. Their result is restricted to the non-paranormal distribution, and our study considers the case where the data follows a scale mixture of normal distributions, as indicated in Assumption \ref{assump-SMN-different-W}.

\begin{lemma}[Limiting spectral distribution] \label{lem:rho_W_LSD} For the high-dimensional factor model \eqref{eq:factor-model}, assume that Assumptions \ref{assump-p-n}  --  \ref{assump-psi} hold, and the ESD of $\bSigma_{\rho}=\Expe\bW_n$ tends to a proper probability measure $H$ as $n\to\infty$. Then, with probability one, 
		both $F^{\brho_n}$ and $F^{\bW_n}$ tend to a non-random probability distribution $F_{c,H}$, the Stieltjes transform $m=m(z) \; (z\in\bbC^{+})$ of which is the unique solution to the equation 
		\begin{equation}\label{eq:LSD_stieltjes_transform}
			m = \int \frac{1}{t(1-c-czm)-z}\dif H(t).
		\end{equation}
\end{lemma}

The following Lemma \ref{lem:G4MT} concerns the limiting behavior of $\bOmega_K(\cdot, \cdot)$ defined in \eqref{eq:def-Omega}, which plays a crucial role in the proof of Theorem \ref{thm:phase-transition}.

\begin{lemma}\label{lem:G4MT}
    Let $\bX=(X_{ij})_{n\times p}=(\bx_1,\ldots,\bx_n)^{\top}$ be an $n\times p$ random matrix. Assume that $\bX$ satisfies Assumption \ref{assump-p-n} and the following assumptions:
    \begin{enumerate}[label={(B\arabic*)}, ref={(B\arabic*)}]
        \item \label{assump-iid-column} The vectors $\{\bx_{\ell}\}_{\ell=1}^n$ are i.i.d., but the entries of each $\bx_{\ell}$ are not necessarily i.i.d.
        \item \label{assump-moment} (Moment condition) For any $i,j,s,t\in[p]$ with $i\neq j\neq s\neq t$, we have $\Expe X_{1i}=0$, $\Expe X_{1i}^2=1$, $\Expe X_{1i}X_{1j}=0$, $\Expe X_{1i}^4=O(1)$, $\Expe X_{1i}^2 X_{1j} X_{1s} =O(p^{-1})$, $\Expe X_{1i} X_{1j} X_{1s} X_{1t}=O(p^{-2})$.
        \item \label{assump-weak-dependency} (Weak dependency) For any $p\times p$ symmetric matrix $\bT$ with bounded spectral norm, we have $\Var(\bx_1^{\top}\bT\bx_1)=o(p^2)$ as $p\to\infty$.
        \item \label{assump-concentration} (Concentration) For any convex $1$-Lipschitz (with respect to the Euclidean norm) function $F$ from $\bbR^p$ to $\bbR$, let $m_F$ denote a median of $F$, we have
        $
        \Prob\bigl(|F(\bx_1)-m_F|>t\bigr) \leqslant C\exp\bigl\{ -c(p)t^2 \bigr\},
        $
        where $C$ and $c(p)$ are independent of $F$, and $C$ is independent of $p$. We allow $c(p)$ to be a constant or to go to zero with $p$ like $p^{-\alpha}$, $0\leqslant \alpha <1$.
    \end{enumerate}
    Moreover, let $\bY=(Y_{ij})_{n\times p}=(\by_1,\ldots,\by_n)^{\top}$ be a random matrix independent of $\bX$, satisfying Assumptions \ref{assump-p-n} and \ref{assump-iid-column}  --  \ref{assump-concentration} with $X_{ij}$ and $\bx_{\ell}$ replaced by $Y_{ij}$ and $\by_{\ell}$, respectively. Then, $\bOmega_K(\lambda, \bX)$ and $\bOmega_K(\lambda, \bY)$ have the same limiting distribution, where $\bOmega_K(\cdot,\cdot)$ is defined in \eqref{eq:def-Omega}.
\end{lemma}
\begin{remark}\label{rmk:distribution_concentration}
    Assumption \ref{assump-concentration} is from \cite{el2009concentration}. In \cite[p.~2386]{el2009concentration}, the author gave some examples of distributions satisfying Assumption \ref{assump-concentration}, such as:
    \begin{itemize}
        \item Gaussian random vectors with covariance matrix $\bSigma_p$ and $c(p)=1/\|\bSigma_p\|_2$ (according to Theorem 2.7 in \cite{ledoux2001concentration}).
        \item Random vectors with i.i.d. entries bounded by $1/\sqrt{c(p)}$ (according to Lemma S1.3).
    \end{itemize}
\end{remark}

\subsection{Proof of Theorem \ref{thm:phase-transition}}\label{sec:proof-phase-transition}
From Lemma \ref{lem:rho_W_spectral_norm}, we
investigate the phase-transition theory of spiked eigenvalues of $\brho_n$ by those of $\bW_n=(3/n)\sum_{i=1}^n\bA_i\bA_i^{\top}$. 
Recall that $\bSigma_{\rho}=\Expe\bW_n$.
Define the spectral decomposition of $\bSigma_{\rho}$ as
$
\bSigma_{\rho} = \bU\bigl(\begin{smallmatrix}
	\bD_1 & \boldsymbol{0}\\
	\boldsymbol{0} & \bD_2
\end{smallmatrix}\bigr)
\bU^{\top},
$
where $\bU$ is a $p\times p$ orthogonal matrix, $\bD_1$ is the diagonal matrix consisting of the $K$ spiked population eigenvalues, and $\bD_2$ is the diagonal matrix consisting of the remaining $p-K$ non-spiked eigenvalues. Let $\widetilde{\bA}_i := \sqrt{3}\bSigma_{\rho}^{-\nicefrac{1}{2}}\bA_i$ denote a transformed version of $\sqrt{3}\bA_i$. It is obvious that $\widetilde{\bA}_i$ is isotropic, that is, $\Cov(\widetilde{\bA}_i)=\bI_p$. By using these notations and the spectral decomposition of $\bSigma_{\rho}$, we have
\begin{equation}\label{eq:characteristic-eq}
    0  = |\lambda\bI_p-\bW_n|
        = \left| \lambda \bI_p - \bU\begin{pmatrix}
            \bD_1^{\nicefrac{1}{2}} & \\ & \bD_2^{\nicefrac{1}{2}}
        \end{pmatrix}\bU^{\top}\widetilde{\bW}_n\bU \begin{pmatrix}
            \bD_1^{\nicefrac{1}{2}} & \\ & \bD_2^{\nicefrac{1}{2}}
        \end{pmatrix} \bU^{\top}\right|,
\end{equation}
where $\widetilde{\bW}_n :=n^{-1}\boldcalA^{\top}\boldcalA$ with $\boldcalA:=(\widetilde{\bA}_1,\ldots,\widetilde{\bA}_n)^{\top}$.
Let $\bQ = \bU^{\top}\widetilde{\bW}_n\bU$ and partition it as
$
\bQ = \Bigl(\begin{smallmatrix}
	\bQ_{11}&\bQ_{12}\\\bQ_{21}&\bQ_{22}
\end{smallmatrix}\Bigr)
=\Bigl(\begin{smallmatrix}
	\bU_1^{\top}\widetilde{\bW}_n\bU_1 & \bU_1^{\top}\widetilde{\bW}_n\bU_2\\
	\bU_2^{\top}\widetilde{\bW}_n\bU_1 & \bU_2^{\top}\widetilde{\bW}_n\bU_2
\end{smallmatrix}\Bigr),
$
where $\bU_1$ is the submatrix formed by the first $K$ columns of $\bU$, and $\bU_2$ is the remaining submatrix. Plugging this identity into \eqref{eq:characteristic-eq} yields that
\begin{align*}
    0 
    & = \left| \lambda \bI_p - \begin{pmatrix}
        \bD_1^{\nicefrac{1}{2}}\bQ_{11}\bD_1^{\nicefrac{1}{2}} & \bD_1^{\nicefrac{1}{2}}\bQ_{12}\bD_2^{\nicefrac{1}{2}} \\ 
        \bD_2^{\nicefrac{1}{2}}\bQ_{21}\bD_1^{\nicefrac{1}{2}} &  \bD_2^{\nicefrac{1}{2}}\bQ_{22}\bD_2^{\nicefrac{1}{2}}
    \end{pmatrix}  \right| \\[0.5em]
    & = \Bigl|\lambda\bI_{p-K}- \bD_2^{\nicefrac{1}{2}}\bQ_{22}\bD_2^{\nicefrac{1}{2}}\Bigr| \\
    &\qquad \times \Bigl|\lambda\bI_K-\bD_1^{\nicefrac{1}{2}}\bQ_{11}\bD_1^{\nicefrac{1}{2}}-\bD_1^{\nicefrac{1}{2}}\bQ_{12}\bD_2^{\nicefrac{1}{2}}(\lambda\bI_{p-K}-\bD_2^{\nicefrac{1}{2}}\bQ_{22}\bD_2^{\nicefrac{1}{2}})^{-1}\bD_2^{\nicefrac{1}{2}}\bQ_{21}\bD_1^{\nicefrac{1}{2}}\Bigr|,
\end{align*}
where the last equality follows from the formula $\det\left(\begin{smallmatrix}\bA&\bB\\\bC&\bD\end{smallmatrix}\right) = \det(\bA-\bB\bD^{-1}\bC)\cdot\det(\bD)$.
Suppose that $\lambda$ is a spiked eigenvalue, then we have $|\lambda \bI_{p-K}- \bD_2^{\nicefrac{1}{2}}\bQ_{22}\bD_2^{\nicefrac{1}{2}}|\neq 0$, and
\begin{align}
    0 
    & =  \Bigl|\lambda\bD_1^{-1}-\bQ_{11}-\bQ_{12}\bD_2^{\nicefrac{1}{2}}(\lambda\bI_{p-K}-\bD_2^{\nicefrac{1}{2}}\bQ_{22}\bD_2^{\nicefrac{1}{2}})^{-1}\bD_2^{\nicefrac{1}{2}}\bQ_{21}\Bigr|\nonumber\\[0.5em]
    & = \Bigl| \lambda\bD_1^{-1} - \frac{1}{n}\bU_1^{\top}\boldcalA^{\top} \Bigl[ \bI_n+\frac{1}{n}\boldcalA\bU_2\bD_2^{\nicefrac{1}{2}}\Bigl(\lambda\bI_{p-K}-\frac{1}{n}\bD_2^{\nicefrac{1}{2}}\bU_2^{\top}\boldcalA^{\top}\boldcalA\bU_2\bD_2^{\nicefrac{1}{2}}\Bigr)^{-1}\bD_2^{\nicefrac{1}{2}}\bU_2^{\top}\boldcalA^{\top} \Bigr]\boldcalA\bU_1\Bigr|\nonumber\\[0.5em]
    & = \Bigl|\lambda\bD_1^{-1}  -\frac{\lambda}{n}  \tr\Bigl\{\Bigl( \lambda\bI_n-\frac{1}{n}\boldcalA\bGamma\boldcalA^{\top} \Bigr)^{-1}\Bigr\}\bI_K + n^{-\nicefrac{1}{2}}\bOmega_K(\lambda,\boldcalA) \Bigr|,\label{eq:characteristic_equation_Omega}
\end{align}
where $\bGamma=\bU_2\bD_2\bU_2^{\top}$ and 
\begin{equation}\label{eq:def-Omega}
\bOmega_K(\lambda, \boldcalA) = \frac{\lambda}{\sqrt{n}} \biggl[ \tr\Bigl\{\Bigl( \lambda\bI_n-\frac{1}{n}\boldcalA\bGamma\boldcalA^{\top} \Bigr)^{-1}\Bigr\}\bI_K-\bU_1^{\top}\boldcalA^{\top}\Bigl( \lambda\bI_n-\frac{1}{n}\boldcalA\bGamma\boldcalA^{\top} \Bigr)^{-1}\boldcalA\bU_1 \biggr].
\end{equation} 
From Lemma \ref{lem:G4MT} and Remark \ref{rmk:distribution_concentration}, if $\widetilde{\bA}_i$ satisfies Assumptions \ref{assump-iid-column}  --  \ref{assump-concentration}, we can replace entries of $\boldcalA$ by the standard Gaussian entries without changing the phase-transition theory of the spiked eigenvalues. Then, our Theorem \ref{thm:phase-transition} follows from Lemma 3.1 and Theorems 4.1 -- 4.2 in \cite{bai2012sample}.

It remains to prove that random vector $\widetilde{\bA}_i$ satisfies Assumptions \ref{assump-iid-column}  --  \ref{assump-concentration}, which shows that our Lemma \ref{lem:G4MT} applies. It is obvious that $\widetilde{\bA}_i$ satisfies Assumption \ref{assump-iid-column}. By using Lemma \ref{lem:Sigma_finite_rank_perturbation}, we conclude that $\liminf_{p\to\infty}\lambda_p(\bSigma_{\rho}^{\nicefrac{1}{2}}) >0$, and thus $\bSigma_{\rho}^{-\nicefrac{1}{2}}$ is bounded in spectral norm. Combining this information and the fact that each component of the random vector $\bA_i$ follows $\mathsf{Uniform}(-1,1)$ distribution, we conclude that each component of the random vector $\widetilde{\bA}_i$ has bounded fourth moment. This, together with Lemma \ref{lem:Sigma_finite_rank_perturbation} and similar calculations in Section S2.3.1 of the supplementary material,
implies that $\widetilde{\bA}_i$ satisfies the moment condition \ref{assump-moment}.
From (S2.16) in the supplementary material 
and the fact that $\|\bSigma_{\rho}^{-\nicefrac{1}{2}}\|_2=O(1)$, we have, for any $p\times p$ symmetry matrix $\bT$ with bounded spectral norm, $\Var(\widetilde{\bA}_i^{\top}\bT\widetilde{\bA}_i)=\Var(3\bA_i^{\top}\bSigma_{\rho}^{-\nicefrac{1}{2}}\bT\bSigma_{\rho}^{-\nicefrac{1}{2}}\bA_i)=o(p^2).$
Hence, $\widetilde{\bA}_i$ satisfies Assumption \ref{assump-weak-dependency}.
By Assumptions \ref{assump-loading} and \ref{assump-psi}, Lemma \ref{lem:Sigma_finite_rank_perturbation}, and the fact that each component of $\bA_i$ follows $\mathsf{Uniform}(-1,1)$, we conclude that each component of $\widetilde{\bA}_i$ is bounded, and thus satisfies the concentration assumption \ref{assump-concentration}, according to Lemma S1.3 in the supplementary material.
Therefore, the random vector $\widetilde{\bA}_i$ satisfies Assumptions \ref{assump-iid-column}  --  \ref{assump-concentration}, which shows that our Lemma \ref{lem:G4MT} applies.
This completes the proof of Theorem \ref{thm:phase-transition}.
    
\subsection{Proof of Theorem \ref{thm:Stieltjes_ratio}}\label{sec:proof-consistency}

From Theorem \ref{thm:phase-transition}, we have,  for any $j\in [K_0]$, $j<\ell\leqslant p$,
$\frac{1}{\{ \lambda_{j}(\brho_n)- \lambda_{\ell}(\brho_n)\}^2}\asymp 1,$
here $x \asymp 1$ means that there exist constants $a$ and $b$ such that $a<x<b$ a.s.. Thus, 
\[
    \widehat{m}_{n,j}'\bigl(\lambda_{j}(\brho_n)\bigr)  
= \frac{1}{p-j} \sum_{\ell=j+1}^p\frac{1}{\bigl\{ \lambda_{j}(\brho_n)- \lambda_{\ell}(\brho_n)\bigr\}^2}\asymp 1,
\] and
\begin{equation}\label{eq:SR_1}
    \frac{	\widehat{m}_{n,j+1}'\bigl(\lambda_{j+1}(\brho_n)\bigr)}{\widehat{m}_{n,j}'\bigl(\lambda_{j}(\brho_n)\bigr)}=O_P(1),\qquad\text{for } j\in [K_0-1].
\end{equation}
From the fluctuation of edge eigenvalues \citep{Peche2009,bao2019tracy-spearman}, we have $\lambda_j(\brho_n)-\lambda_{j+1}(\brho_n)=O_P(p^{-\nicefrac{2}{3}})$ for $K_0+1\leqslant j\leqslant K_{\max}$. Thus, for $K_0+1\leqslant j\leqslant K_{\max}$, we get
\[
    \widehat{m}_{n,j}'\bigl(\lambda_{j}(\brho_n)\bigr)  
    = \frac{1}{p-j} \sum_{\ell=j+1}^p\frac{1}{\bigl\{ \lambda_{j}(\brho_n)- \lambda_{\ell}(\brho_n)\bigr\}^2}
    =O_P(p^{\nicefrac{1}{3}}).
\]
Therefore, 
\begin{equation}\label{eq:SR_2}
    \frac{	\widehat{m}_{n,K_0+1}'\bigl(\lambda_{K_0+1}(\brho_n)\bigr)}{\widehat{m}_{n,K_0}'\bigl(\lambda_{K_0}(\brho_n)\bigr)} \to \infty. 
\end{equation}	
Moreover, for $K_0+1\leqslant j\leqslant K_{\max}$, we have
\begin{align}
    \frac{	\widehat{m}_{n,j+1}'\bigl(\lambda_{j+1}(\brho_n)\bigr)}{\widehat{m}_{n,j}'\bigl(\lambda_{j}(\brho_n)\bigr)} 
    = \frac{p-j}{p-j-1}\frac{\sum_{\ell=j+2}^p\bigl\{ \lambda_{j+1}(\brho_n)- \lambda_{\ell}(\brho_n)\bigr\}^{-2}}{\sum_{\ell=j+1}^p\bigl\{ \lambda_{j}(\brho_n)- \lambda_{\ell}(\brho_n)\bigr\}^{-2}} \asymp 1.\label{eq:SR_3}
\end{align}
By combining \eqref{eq:SR_1}  --  \eqref{eq:SR_3}, we obtain that $\widehat{K}_{\SR}$ is consistent. Moreover, if $\widehat{m}'_{n,j}(x)$ is replaced by $\widetilde{m}'_{n,j}(x)$, the derivations above remain valid, thereby ensuring the consistency of $\widetilde{K}_{\SR}$ in \eqref{eq:ktilde} as well.

\bigskip
\begin{center}
{\large\bf SUPPLEMENTARY MATERIAL}
\end{center}

This supplementary material contains some auxiliary lemmas and the technical proofs of Lemmas \ref{lem:rho_W_spectral_norm}, \ref{lem:Sigma_finite_rank_perturbation}, \ref{lem:rho_W_LSD}, \ref{lem:G4MT}, S1.6, and S1.7.






\bibliographystyle{apalike}
\bibliography{ref}

\end{document}